\documentclass[12pt]{article}
\usepackage{mathrsfs}
\usepackage{amsmath,amssymb}
\usepackage{epsfig}
\usepackage{relsize}
\usepackage{bbm}
\usepackage{graphicx}
\usepackage{float}
\usepackage{color}
\usepackage{enumitem}

\usepackage{sidecap}

\def\({\left(}
\def\){\right)}

\def\sl(2){\alg{sl}(2)}

\def\be{\begin{equation}}
\def\ee{\end{equation}}

\newcommand{\bea}{\begin{eqnarray}}
\newcommand{\eea}{\end{eqnarray}}

\newcommand{\bei}{\begin{itemize}}
\newcommand{\eei}{\end{itemize}}

\newcommand{\bee}{\begin{enumerate}}
\newcommand{\eee}{\end{enumerate}}

\def\a {\alpha}

\def\s {\sigma}

\def\g {\gamma}

\def\p{\phi}
\def\la{\label}
\def\e{\epsilon}
\def\ov{\over}

\def\vp{\varphi}

\newcommand{\alg}[1]{\mathfrak{#1}}
\newcommand{\su}{\alg{su}}

\newcommand{\bem}{\left (\begin{matrix}}
\newcommand{\eem}{\end{matrix} \right )}

\def\ka{\kappa}

\definecolor{grey}{rgb}{0.4,0.4,0.5}
\definecolor{darkgreen}{rgb}{0,0.5,0}
\definecolor{darkred}{rgb}{0.6,0.0,0}
\definecolor{lightbrown}{rgb}{1,0.9,0.8}
\definecolor{brown}{rgb}{0.6,0.3,0.3}
\definecolor{darkblue}{rgb}{0,0,0.8}
\definecolor{darkmagenta}{rgb}{0.5,0,0.5}

\newcommand{\Der}[2]{{{\rm d} #1 \ov {\rm d} #2}}

\def\Reals{{\mathbb R}}

\def\I{{\cal I}}

\def\Gadd{{\mathfrak a}}
\def\Grem{{\mathfrak r}}

\def\Radd{{\rm p}}
\def\Rrem{{\rm h}}

\def\Tadd{{\rm a}}
\def\Trem{{\rm r}}

\def\Nadd{{N^\Tadd}}
\def\Nrem{{N^\Trem}}

\def\barew{{\rm w}}

\def\cm{\mathbf c}
\def\cd{{\mathbf c}^\dagger}
\def\n{\mathbf n}

\def\n{\mathbf n}

\def\hstar{\,\hat{\star}\,}
\def\cstar{\,\check{\star}\,}

\def\uh{ \mathfrak u}

\def\E{\mathcal E}

\def\ve{{\mbox{\large$e$}}}

\def\brho{\bar\rho}

\def\cK{{\cal K}}

\newcommand{\bean}{\begin{eqnarray*}}
\newcommand{\eean}{\end{eqnarray*}}

\usepackage{hyperref}
\hypersetup{
plainpages=false
}

\begin{document}



\null\vskip-40pt
 \vskip-5pt \hfill
\vskip-5pt \hfill {\tt\footnotesize
TCD-MATH-12-11}
 \vskip-5pt \hfill {\tt\footnotesize
HMI-12-04}

\vskip 1cm \vskip0.2truecm
\begin{center}
\begin{center}
\vskip 0.8truecm 
{\Large\bf Excited states in Bethe ansatz solvable models 
 
 \vspace{0.3cm}
 
and  the dressing of spin and charge}
\end{center}

\renewcommand{\thefootnote}{\fnsymbol{footnote}}

\vskip 0.9truecm
Eoin Quinn\footnote[1]{emails: 
epquinn@gmail.com, frolovs@maths.tcd.ie}\ and Sergey Frolov\footnote[2]{Correspondent fellow at
Steklov Mathematical Institute, Moscow.} 
 \\
\vskip 0.5cm

{\it School of Mathematics and Hamilton Mathematics Institute, \\
Trinity College, Dublin 2,
Ireland}

\end{center}
\vskip 1cm \noindent\centerline{\bf Abstract} \vskip 0.2cm

A general formalism for the study of excitations above equilibrium in Bethe ansatz solvable models is presented. 
Nonzero temperature expressions for dressed energy, momentum,  spin and charge are obtained.
The zero temperature excitations of the Hubbard-Shastry models are examined in detail, and special attention is paid to the dressing of spin and charge of excited quasi-particles. These are in general momentum dependent and are only spin-charge separated  when the ground state is half-filled and has zero magnetisation.

\newpage

\tableofcontents

\renewcommand{\thefootnote}{\arabic{footnote}}
\setcounter{footnote}{0}

\numberwithin{equation}{section}



\section{Introduction}

The quest to understand strongly correlated electrons in low-dimensional systems represents an important frontier in the field of condensed matter physics \cite{MIT, LNW06}. Such materials exhibit exotic behaviour that cannot be understood from a non-interacting picture. Exactly solvable models are an invaluable tool for gaining access to the non-perturbative physics at play.

Two recently introduced integrable models, the Hubbard-Shastry A- and B-models \cite{FQ11}, hold great  promise in this direction. 
They describe  electrons interacting on a lattice, similar to the well known Hubbard model \cite{Hubbard, bookH}, but with extra interactions such as spin exchange, pair hopping and nearest neighbour Coulomb interaction.
Their equilibrium properties were examined in \cite{FQ11}  by means of their exact Bethe ansatz solution. It was found that the A-model exhibits  itinerant ferromagnetism, while the B-model is a Mott insulator of paired electrons and so provides a promising approach to the study of unconventional superconductivity  given that such physics is expected to be captured by an effective single-band model \cite{Anderson, ZhRice}.

The aim of this paper is to examine the excitations above equilibrium of these two models  in one-dimension.  To achieve this goal a general formalism of excitations in Bethe ansatz solvable models is pursued. This generalises and extends the methods used in  \cite{YY69}-\cite{TSO}. Let us briefly outline some of the advancements. Firstly, we consider individual particle and hole excitations and obtain explicit non-zero temperature expressions for dressed quantities such as energy, momentum, spin and charge. 
Our expressions include an important contribution that did not appear in previous studies which were restricted to particle-hole excitations \cite{KBI93}. Secondly, we examine in some detail the dressing of spin and charge and provide a formula for the induced charge of the system that results from an excitation. Thirdly, we consider models with Bethe strings and extend to such models the formalism for excitations presented in e.g. \cite{KBI93} for the Bose gas.  Fourthly, when considering excitations at zero temperature we overcome the need to explicitly deal with mode numbers \cite{Hult,dCP}, and relate possible restrictions on allowed excitations to properties of the kernels appearing in the Thermodynamic Bethe Ansatz equations.

The dressing of spin and charge is well established and in integrable models it dates back to \cite{FadTak}. The zero temperature long-range physics of many one-dimensional models of electrons is captured by the Luttinger liquid, wherein spin-charge separation is exhibited, see e.g. \cite{G1D}. The low-lying wave-like excitations  carry either spin or charge and propagate at different velocities. Let us stress however that individual quasi-particle excitations need not exhibit spin-charge separation, even at zero temperature. In general these carry both spin and charge, the magnitude of which  varies with the momentum of the excited quasi-particle. This is compatible with the Luttinger liquid behaviour as wave-like excitations are composed of infinitely many quasi-particles carrying infinitesimal energies, and in this limit spin-charge separation may be seen. 
We would like to suggest that such non-trivial dressing of spin and charge may account for some of the anomalous behaviour observed in strongly correlated materials. It would be interesting to consider issues such as the unusual temperature dependence  of resistivity, the Hall coefficient, and spin excitations in this context, see e.g. \cite{RlinearT,spinstudies}.

The paper is naturally split into two parts. In the first the formalism for the study of excitations above equilibrium is presented and 
in the second part of the paper the zero temperature limit of this formalism is used to examine excitations over the ground states of the Hubbard-Shastry A- and B-models.

 In  Bethe ansatz solvable models  the complexity of calculating the spectrum is reduced to the solution of the Bethe equations, a set of  polynomials whose degree scales linearly with the length of the system. In the thermodynamic limit the roots of these equations, in general complex, align into strings on the complex plane known as Bethe strings. It is often possible to make a string hypothesis, which identifies all possible Bethe strings, and in this paper we restrict our attention to models for which this is so. 

There are advantages to working directly with the Bethe strings in the thermodynamic limit. 
The Bethe strings are parametrised by a rapidity variable that is real whereas, as we have mentioned, their constituents (the roots of the Bethe equations) generally take complex values. The original Bethe equations can be fused together to give Bethe equations directly for the strings. Moreover, the Bethe strings give a physical picture: the  different types of strings can be understood as the particle content of the model. 
The free energy can thus be written and its minimisation gives access to the equilibrium state. In this way one goes from the full information about the spectrum that the Bethe equations provide to the physically interesting region. 

 To examine excitations above equilibrium  it is also natural to work directly with the Bethe strings  and this is the approach taken in the formalism developed in section \ref{formalism}.  Excitations of a finite number of Bethe strings are considered and formulae for the dressed energy, charge and spin of the excited strings are obtained, as well as the scattering phases for the excited roots which are non-trivial and include scattering with the equilibrium state.  A summary of these results is provided in section \ref{summary}.

In the zero temperature limit the equilibrium state becomes the ground state.
This limit is in general smooth, with the special exception of when some strings are at half-filling. By a half-filled string we mean one for which there are no holes in the ground state. 
In such cases there may exist restrictions on allowed excitations over the ground state.  This can be seen as a manifestation of a zero temperature phase transition.

Now we turn to our study of the excitations of the Hubbard-Shastry models, which comprises the second part of this paper. 
Attention is focused  on the zero-temperature regime as the essential features can be observed there. General features of finite temperature behaviour are discussed throughout the presentation of section \ref{formalism}.
A detailed analysis of the finite temperature excitations requires the solution of various infinite  sets of coupled non-linear integral equations and is not pursued here.

The A-model describes an itinerant ferromagnet. It has a gound state that is spin-polarised for all fillings. At half-filling the behaviour is similar to that of the ferromagnetic Heisenberg XXX spin chain. The low-lying excitations  are magnons and their bound states and they disperse quadratically. In the presence of a non-zero magnetic field they become gapped.  In addition there is an electron-like excitation. This is gapped at half-filling but becomes gapless with linear dispersion away from half-filling. The spin of the electron-like excitation is aligned with that of the spin-polarised ground state. An electron-like excitation of opposite spin should be regarded as a compound excitation of this aligned electron-like excitation and a magnon. The excitations that behave as magnons at half-filling retain their quadratic dispersion away from half-filling but here get dressed with some charge in addition to their spin. 
Thus the spectrum of low-lying excitations cannot be linearised at any filling in zero magnetic field and the system cannot be approximated by a Luttinger liquid, as is evidenced in the lack of spin-charge separation. In the presence of a magnetic field the dressed magnons get gapped and there are then no low-lying excitations that decrease the magnetisation of the ground state without decreasing the filling.

The B-model describes a Mott insulator of spin-singlets.
At half-filling and in the absence of a magnetic field the model is in an insulating phase and the magnetisation is zero.  Here the excitations are spin-charge separated for all momenta, they are scattering states of gapless spinons and gapped holons and the picture is similar to that of the Hubbard model \cite{bookH}. The difference is that here the spinons are dressed electrons and the holons are dressed spin singlets (paired electrons) whereas in the Hubbard model the spinons are dressed magnons and it is the holons that are dressed electrons. If one goes away from half-filling or introduces a magnetic field then the spin and charge are no longer separated and become rapidity dependent. In particular away from half-filling in the absence of a magnetic field the dressed spin singlets remain pure charge and become gapless while the electrons are also gapless but get dressed with a charge in addition to their spin, and thus the excitations are not spin-charge separated.  The dressed charge of an excited electron does go to zero however as its dressed energy goes to zero  and so these excitations are compatible with those of a Luttinger liquid which is expected in the continuum limit. Let us conclude by remarking  that the dressed electrons display an ``hourglass''  dispersion  and  that  away from half-filling and in the absence of a magnetic field  it can be clearly seen that the spin singlets are held together by spin-spin interactions.

\section{Excited states} \la{formalism}
 
In this section we develop a general formalism to study excited states of  Bethe ansatz solvable models whose thermodynamics are described by various configurations of Bethe strings \cite{Takbook}. The dressed energy, momentum and charge, and also the  phase shifts of the excitations, are expressed through the densities of the equilibrium state. The limit of zero temperature is examined in section \ref{zeroT} and the formalism is summarised in section \ref{summary}. For simplicity and clarity of presentation we restrict ourselves to parity invariant rational models.

\subsection{Bethe equations for strings}

Let us assume that in the thermodynamic limit 
every solution of the Bethe equations  for the model under consideration  corresponds to a particular configuration of Bethe strings.  
Then for large spin chain length $L$ the Bethe equations can be written for string configurations
\be\la{BE1}
 (-1)^{\varphi_\a}=e^{iLp_\a(v_{\a,k}) }\,\prod_\beta\prod_{n=1}^{N_\beta}\, S_{\a\beta}(v_{\a,k},v_{\beta,n})\,.
\ee
We use indices $\a$, $\beta$, $\g$ and so on to distinguish between different strings. 
Here $\varphi_\a$ is a constant, unimportant for our consideration, which appears in particular as we have not excluded self-scattering from the product on the right hand side. For periodic boundary conditions and in the absence of a twist, $\varphi_\a=0$ if $S_{\a\a}(0)=1$ and $\varphi_\a=1$ if $S_{\a\a}(0)=-1$. The strings are parametrised by a rapidity variable for which the scattering matrices are of a difference form: $S_{\a\beta}(v_{\a,k},v_{\beta,n})=S_{\a\beta}(v_{\a,k}-v_{\beta,n})$.
 For strings which do not carry momentum there is no term $e^{iLp_\a}$ in the Bethe equations, and by convention we set $p_\a=0$ for such strings. 

Let us briefly review the analysis that yields the equilibrium properties of the model. This allows us to introduce many useful formulae, along with our conventions. 
The counting function is constructed by taking the logarithm of the Bethe equations \eqref{BE1}
\be\la{za1}
L \,\s_\a z_\a(v)=\pi{\varphi_\a}+ {L\, p_\a(v)}+\sum_\beta\sum_{n=1}^{N_\beta}\,\p_{\a\beta}(v,v_{\beta,n})\,,
\ee
where
\be\la{pab}
\p_{\a\beta}\equiv{1\ov i}\log S_{\a\beta}
\ee
is the phase of the S-matrix. 
These functions allow one to enumerate the Bethe strings as $L \, z_\a\ov 2\pi$ evaluates to an integer  on an $\a$-string Bethe root, which we refer to as the mode number of the particle. For a given root $v_{\a,k}$ we denote the corresponding mode number as $I_{\a,k}\equiv{L\ov 2\pi}z_\a(v_{\a,k})$. Moreover the counting function $Lz_\a/2\pi$ may evaluate to an integer for a momentum which is not that of a particle of type $\a$, and such mode numbers correspond to holes.  
Note the appearance of $\s_\a$ in the definition of the counting function. 
For momentum carrying strings $\s_\a=\mbox{sign}(\Der{p_\a}{v})$ and this guarantees that the counting function is an increasing function of $v$. 
For an $\a$-string which does not carry momentum $\s_\a$ is determined by requiring the counting function to be increasing. Let us remark that to fully define the counting function \eqref{za1} it is necessary to specify that branch of the logarithm. This does not affect the study of equilibrium properties, and of course should not affect the physics, but it will be important for our study of excitations and we will return to this issue  when we begin to discuss them in section \ref{exstates}.

Taking the thermodynamic limit $L\to\infty$ with $N_\a/L$ fixed, one gets  equations for the densities of particles and holes
\be\la{rbr}
\rho_{\a} + {\bar{\rho}}_{\a} = \frac{1}{2 \pi} \left|\Der{p_\a}{v}\right|+ K_{\a\beta}\star\rho_{\beta} \,,
\ee
where $L \rho_{\a} {\rm d}v$ (respectively $L \brho_{\a} {\rm d}v$) is the number of integers corresponding to particles (respectively holes) that the counting function $Lz_\a/2\pi$ evaluates to over a range ${\rm d}v$.
Repeated indices are summed over and $\star$ denotes convolution (see appendix \ref{conventions} for the precise definition) over the domain of the rapidity of the appropriate string, which we denote by $\I_\a$. The kernels $K_{\a\beta}$ are defined by
\be\la{sKer}
K_{\a\beta}=\s_\a \cK_{\a\beta} \,,\quad \cK_{\a\beta}(v)={1\ov 2\pi i}{d\ov dv}\log S_{\a\beta}(v) \,.
\ee
Since the counting functions are all defined to be increasing functions it follows that the densities of all particles and holes in the  equations \eqref{rbr} are positive.

The equations for the densities can be used to determine the total number of  particles and holes for each type of string, which we call the range of mode numbers.  Indeed, integrating equations \eqref{rbr} and multiplying by $L$ one gets the range
 \be\la{range1}
N_{\a} + \bar{N}_{\a} =L\,{\Delta p_\a\ov 2\pi} + k_{\a\beta}N_{\beta}  \,,
\ee
where  $ \bar{N}_{\a}$ is the number of holes of type $\a$ in the state, and $\Delta p_\a=1\star |\Der{p_\a}{v}|$ for momentum carrying strings and $\Delta p_\a=0$ for  strings that do not carry momentum. The $ k_{\a\beta}$, defined as $ k_{\a\beta}\equiv 1\star K_{\a\beta}$, are constants. Note that all roots of the Bethe equations are counted in the range, including those which correspond to $v=\infty$.  Furthermore since $\bar{N}_{\a}\ge 0$ eqs. \eqref{range1} imply the following selection rules 
 \be\la{rule1}
N_{\a} \le L\,{\Delta p_\a\ov 2\pi}+ k_{\a\beta}N_{\beta}  \,,
\ee
which restrict the allowed $N_\a$ appearing in the Bethe equations.

\subsection{Equilibrium  state}

The equilibrium state follows from the minimisation of the free energy
\be\la{freng}
f=\ve_\a \star \rho_\a - T s\,.
\ee
Here $\ve_\a$ denotes the energy carried by an $\a$-string. Note that in addition to describing possible dispersion this will also depend on the chemical potentials that appear in the model. The entropy per site is denoted by $s$ and is given by
\be
s=\sum_\a 1\star {\Delta s}(\rho_\a,\bar\rho_\a)\,,\quad
  {\Delta s}(\rho,\bar\rho) = \rho \log\big(1+{\bar\rho\ov\rho}\big)+ \bar\rho \log\big(1+{\rho\ov\bar\rho}\big)\,.
  \ee
Minimising the free energy subject to the equations for densities \eqref{rbr} yields
\be
f=-{ T\ov 2\pi} \left|\Der{p_\a}{v}\right|\star \log\big(1+{1\ov Y_\a}\big)\,,
\ee
where the functions $Y_\a$ are determined by  the set of thermodynamic Bethe ansatz (TBA) equations
\be\la{tbae}
\log Y_\a = {\ve_\a\ov T} -\log\big(1+{1\ov Y_\beta}\big)\star K_{\beta\a}\,.
\ee
These should be regarded as the conditions for equilibrium. 
The $Y_\a$ are related to the densities of particles and holes as $Y_\a={\bar\rho_\a \ov \rho_\a}$ and allow one to solve \eqref{rbr} for the densities at equilibrium.
It is useful to also introduce the pseudo-energy
\be
\e_\a = T \log Y_\a\,.
\ee
These are well defined in the zero temperature limit whereas the $Y_\a$ are not. Both quantities $Y_\a$ and $\e_\a$ will be useful and we will use them interchangeably. 

At finite temperature  the equilibrium state is a mixture of 
infinitely many states contributing equally to the partition function. 
We assume for definiteness that the spin chain length $L$ is very large,  and choose any of the states which in the thermodynamic limit contribute to
the equilibrium state as a {\it reference} state. In the zero temperature limit it becomes the ground state of the model. Let us say then that reference state consists of $N_\a$ $\a$-strings with rapidities $v_{\a,k}$, $k=1,\dots,N_\a$.

\subsection{Excited states}\la{exstates}

Now we wish to study excitations about the equilibrium state outlined above. We restrict our attention to excitations where the numbers of excited roots are much smaller than the numbers of particles in the equilibrium state. 

Let us first return to a point that we skipped above, the choice of the branch for the counting function \eqref{za1}. The choice of branch affects the formalism one obtains for the excitations and we present here what we found to be an optimal choice. 
 In particular we find that in order to obtain a reasonable expression for the dressed momentum of a string it is necessary to keep track of the branch of each $\log S_{\a\beta}$ term in eq.\eqref{za1}. Moreover, our choice of branch is guided by the behaviour of the pseudo-energies for the equilibrium state. We choose the branch of  $\log S_{\a\beta}$ and the range of  momentum $p_\a$ so that the counting function is continuous about  the minimum of $\e_\a$.

 Let us be specific about the strings we will consider. For strings with rapidity variable defined on $\Reals$ we assume that the pseudo-energy is even, and monotonic on the interval $(0,\infty)$. Then there are two cases, and we give explicit expressions for the phase in each
\bei
\item Type 1: $\e_\a(v)$ has a minimum at $v=0$, and is increasing on the interval $(0,\infty)$,
\be\la{case1}
{1\ov i}\log S_{\a\beta}(v,t)=2 \pi b_{\a\beta} + \pi c_{\a\beta} + \Theta_{\a\beta}(v-t)=\p_{\a\beta}(v,t)\,,
\ee
and the range of $p_\a(v)$ is chosen so that it is continuous for $v$ along $(-\infty,\infty)$. 
\item Type 2:  $\e_\a(v)$ has a minimum at $v=\pm\infty$, and is  decreasing on the interval $(0,\infty)$,
\be\la{case2}
{1\ov i}\log S_{\a\beta}(v,t)=2 \pi b_{\a\beta}-\pi c_{\a\beta}\,\mbox{sign}(v) + \Theta_{\a\beta}(v-t)=\p_{\a\beta}(v,t)\,,
\ee
and the range of $p_\a(v)$ is chosen so that $p_\a(-\infty)=p_\a(+\infty)$, and it is discontinuous only at $v=0$.
\eei 
Here 
\be\la{Thdef}
\Theta_{\a\beta}(v)=2\pi\int_{0}^v {\rm d}t \,\cK_{\a\beta}(t)\,,\quad
\ka_{\a\beta}={1\ov \pi}\Theta_{\a\beta}(\infty)=1\star \cK_{\a\beta}\,,
\ee 
$b_{\a\beta}$ is an integer capturing the freedom in the choice of branch, and the $c_{\a\beta}$ are defined such that $c_{\a\beta}-\ka_{\a\beta}$ is as close to zero as possible subject to the constraint that $c_{\a\beta}$ is an even integer if $S_{\a\beta}(0)=1$ and $c_{\a\beta}$ is an odd integer if $S_{\a\beta}(0)=-1$. 
Note that if $S_{\a\beta}(\pm\infty)=1$ (which is the case for rational S-matrices)  then $c_{\a\beta}=\ka_{\a\beta}$. 
For type 1 strings  $\p_{\a\beta}(\pm\infty,t)=\pi(c_{\a\beta}\pm \ka_{\a\beta})$ and the range of $\p_{\a\beta}$ is  $\p_{\a\beta}(+\infty,t)-\p_{\a\beta}(-\infty,t)=2\pi \ka_{\a\beta}$. 
For type 2 strings it is worth stressing that the scattering phase $\p_{\a\beta}$ is no longer of a difference form with this choice of the branch of  $\log S_{\a\beta}$.  It is mildly broken so that,  for models with $c_{\a\beta}=\ka_{\a\beta}$, the counting function and scattering phases would be continuous everywhere but at $v=0$, the maximum of $\e_\a$. The jump discontinuity of the scattering phase at $v=0$ is equal to $\p_{\a\beta}(-0,t)-\p_{\a\beta}(+0,t)=2\pi c_{\a\beta}$, which is equal to $2\pi \ka_{\a\beta}$ if  $S_{\a\beta}(\pm\infty)=1$.

These two cases capture the behaviour of most strings of Bethe ansatz solvable models. For strings that are not captured modifications of the subsequent analysis will have to be made\footnote{ For instance there are  models where the counting function is not always monotonic. So long as it oscillates only a finite number of times however this should not affect the description of excitations above equilibrium.}. In particular, the $y$-particles of the Hubbard-Shastry models must be considered separately because their rapidity variable is not defined on $\Reals$. We deal with this important case in appendix \ref{HSy}.
  Let us remark that although the above definition for type 2 strings \eqref{case2} may at first sight appear to be overkill it is necessary in order to correctly identify the nature of excitations  of such strings, for example the hourglass-like dispersion seen in the Hubbard-Shastry B-model discussed in section \ref{Bmodel}.

Now consider a general excited state. Let us say that it  consists of $\tilde N_\a$  $\a$-strings with rapidities $\tilde v_{\a,k}\,,\ k=1\,,\ldots \,,N_\a$. 
 The rapidities satisfy the same Bethe equations \eqref{BE1}
\be\la{BE2}
  (-1)^{\varphi_\a}=e^{iLp_\a(\tilde v_{\a,k})}\,\prod_\beta\prod_{n=1}^{\tilde N_\beta}\, S_{\a\beta}(\tilde v_{\a,k}-\tilde v_{\beta,n})\,.
\ee
The rapidities $\tilde v_{\a,k}$ of the excited state can be divided into two groups. The first group consists of rapidities with mode numbers  $\tilde I_{\a,k}$ which coincide with some of the mode numbers of the particles of the reference state. They
are close to the corresponding rapidities of the reference state, that is, the difference between the rapidities with the same mode number is of order $1/L$. We denote these rapidities as 
$v_{\a,k}'\,,\ k=1\,,\ldots \,,N_\a'$. 
 The second group consists of the remaining rapidities, those which have mode numbers not coinciding with any mode number  of the particles of the reference state.
There are 
$N^\Tadd_\a=\tilde N_\a -N_\a' $ of strings of type $\a$ in this group. These rapidities will be denoted as $\tilde v_{\Gadd_j}$. The reference state also contains  strings with mode numbers different from any $\tilde I_{\a,k}$, the mode numbers of the particles of the excited state. These correspond to holes of the excited state and their rapidities will be denoted as $\tilde v_{\Grem_j}$. 
For a string of type $\a$ there are $N^\Trem_\a=N_\a -N_\a'$ 
of them. One can think about the excited state as being obtained  by adding $N^{\Tadd}_\a$ strings to, and removing $N^{\Trem}_\a$ strings from, the reference state. 
Let  $\Nadd=\sum_\a N^{\Tadd}_\a$ be the total number of strings added to the reference state, and let $\Nrem=\sum_\a N^{\Trem}_\a$ be  the total number of  strings removed from the reference state.
Thus the equations \eqref{BE2} can be rewritten in the form
\bea\la{BE3}
&& (-1)^{\varphi_\a}=e^{iLp_\a(v_{\a,k}')}\,{\prod_j^{\Nadd}S_{\a\Gadd_j}(v_{\a,k}'-\tilde v_{\Gadd_j})\ov \prod_j^{\Nrem}S_{\a\Grem_j}(v_{\a,k}'-\tilde v_{\Grem_j})}\,\prod_\beta\prod_{n=1}^{N_\beta}\, S_{\a\beta}(v_{\a,k}'-v_{\beta,n}')\,,\\
\la{BE4}
&& (-1)^{\varphi_{\Gadd_k}}=e^{iLp_\a(\tilde v_{\Gadd_k})}\,{\prod_j^{\Nadd}S_{\Gadd_k\Gadd_j}(\tilde v_{\Gadd_k}-\tilde v_{\Gadd_j})\ov \prod_j^{\Nrem}S_{\Gadd_k\Grem_j}(\tilde v_{\Gadd_k}-\tilde v_{\Grem_j})}\,\prod_\beta\prod_{n=1}^{N_\beta}\, S_{\Gadd_k\beta}(\tilde v_{\Gadd_k}-v_{\beta,n}')\,,
\eea
where the product
 $\prod_\beta\prod_{n=1}^{N_\beta}\, S_{\a\beta}(\cdot ,v_{\beta,n}')$ includes the product $\prod_j^{\Nrem}S_{\a\Grem_j}(\cdot,\tilde v_{\Grem_j})$, that is it is equal to $\prod_\beta\prod_{n=1}^{N_\beta'}\, S_{\a\beta}(\cdot ,v_{\beta,n}')\prod_j^{\Nrem}S_{\a\Grem_j}(\cdot,\tilde v_{\Grem_j})$ and reduces to $\prod_\beta\prod_{n=1}^{N_\beta}\, S_{\a\beta}(\cdot,v_{\beta,n})$ in the thermodynamic limit. 
  
Here we have made an implicit assumption that the mode numbers 
of every string in the reference state are also mode numbers of the excited state. This is justified at non-zero temperature as one can always choose the reference state so that this is the case\footnote{Indeed the densities $\rho(v)$, $\brho(v)$ are related to the numbers of particles and holes with rapidity in an interval d$v$ about $v$. At non-zero temperature both the densities of particles and holes are non-trivial for all $v$ and the particles and holes within each interval d$v$ can be rearranged. Thus one can always choose a reference state at non-zero temperature such that a finite number of particles and holes have required mode numbers.}. At zero temperature however, specifically at half-filling, it may happen
 that the mode number 
of a string in the reference state is not a mode number of the excited state, either due to a change in the range of mode numbers or an overall shift of the range of the counting function. For example if the range is decreased for an excitation and all mode numbers correspond to particles then some strings are necessarily removed, and moreover there is no excited state rapidity one can assign to them. Such situations require special care and are discussed in section \ref{zeroT}.

We relate the rapidities of the reference state Bethe roots to those with corresponding mode number in the excited state through
\be\la{defxi}
\tilde v_{\a,k} - v_{\a,k} = {\s_\a\ov 2\pi}{\zeta_{\a}(v_{\a,k})\ov L}\,,
\ee
where we have introduced $\zeta_{\a}$ which are of order 1. 
It is possible to obtain a closed equation for $\zeta_\a$ by subtracting the logarithm of the Bethe equations of the ground state \eqref{BE3} from those of the excited state \eqref{BE1} for Bethe roots with the same mode number. Expanding  \eqref{BE3}  and taking the thermodynamic limit one obtains 
\be\la{xir}
\zeta_\a(\rho_{\a} + {\brho}_{\a} )=\zeta_\beta\rho_{\beta}\star K_{\beta\a}-\p_{\a{\Tadd}}+\p_{\a{\Trem}} \,,
\ee
with the help of the equation for densities \eqref{rbr}. Here we have taken into account that 
\be\la{dlogS2}
S_{\a\beta}(v) S_{\beta\a}(-v)=1\ \Rightarrow\  \frac{\s_\beta}{2 \pi i} \, \frac{d}{dt} \log S_{\a\beta}(v-t)  = -K_{\beta\a}(t-v)\,,
\ee
and introduced the notation
\be\la{notar}
X_{\Tadd}\equiv  \sum_{j=1}^{\Nadd} X_{\Gadd_j}(\tilde v_{\Gadd_j})\,,\quad X_{\Trem}\equiv  \sum_{j=1}^{\Nrem} X_{\Grem_j}(v_{\Grem_j})\,,
\ee
for any quantity $X_\a$.
The function $F_\a=-\zeta_\a (\rho_{\a} + {\brho}_{\a} )$, which we will refer to as the shift function \cite{Korepin79,KBI93}, is an important object and it satisfies the following closed equation 
\be\la{eqFderv}
F_\a={F_\beta\ov 1+Y_\beta}\star K_{\beta\a} +\p_{\a{\Tadd}}-\p_{\a{\Trem}}\,.
\ee
Another form of this equation which will prove useful is
\be\la{xir2}
\zeta_\a\rho_{\a}=\zeta_\beta\rho_{\beta}\star {K_{\beta\a}\ov 1+Y_\a}- {\p_{\a{\Tadd}}-\p_{\a{\Trem}}\ov 1+Y_\a}\,.
\ee

\subsubsection*{Energy}

The change in the energy  of the excited state from the equilibrium state is given by
\be\la{changeE}
\Delta E = \sum_{\a}\Big(\sum_{k=1}^{\tilde N_\a}\, \ve_\a(\tilde v_{\a,k}) - \sum_{k=1}^{N_\a}\,\ve_\a(v_{\a,k})\Big) \to   \ve_\Tadd-\ve_\Trem +{\s_\a \ov 2\pi}\zeta_\a\rho_\a\star \ve_\a'\,,
\ee
where $ \ve_\a' = \Der{}{v}\ve_\a(v)$, and summation over $\a$ is assumed, and we have used the notation \eqref{notar}.
Differentiating the TBA equations \eqref{tbae} and integrating by parts
one gets
\be\la{Epr}
{\ve_\a'\ov T} = (\log Y_\a)' -K_{\a\beta}\star {(\log Y_\beta)'\ov 1+{Y_\beta}} \,,
\ee
and substituting into \eqref{changeE} gives
\be\la{changeE2}
\Delta E =  \ve_{\Tadd}-\ve_{\Trem}+ T{\s_\a \ov 2\pi}\zeta_\a\rho_\a\star \big( (\log Y_\a)' -{K_{\a\beta} \ov 1+{Y_\beta}}\star (\log Y_\beta)'\big) \,,
\ee
Dependence of $\Delta E$ on $\zeta_\a$ can be eliminated through eq. \eqref{xir2} yielding 
\be
\Delta E =  \ve_{\Tadd}-\ve_{\Trem}+{T\ov 2\pi}\,\s_\a\big(\log (1+{1\ov Y_\a})\big)' \star( \p_{\a{\Tadd}}-\p_{\a{\Trem}})  \,.
\ee
Then integrating by parts gives \footnote{Let us remark that the jump discontinuity of $\p_{\a\beta}$ for type 2 strings is $2\pi c_{\a\beta}$, which is equal to $2\pi k_{\a\beta}$ only if $S_{\a\beta}(\pm\infty)=1$.}
\be\la{changeE3}
\Delta E = \ve_{\Tadd}-\ve_{\Trem}-T\, \log (1+{1\ov Y_\a})\star \big(K_{\a{\Tadd}}- K_{\a{\Trem}}\big) +T\log (1+{1\ov Y_\a^{\rm max}})( k_{\a{\Tadd}}-k_{\a{\Trem}})  \,,
\ee
where $Y_\a^{\rm max} = Y_\a(v^{\rm max})$ is equal to $Y_\a$ evaluated at the value of $v$ corresponding to the maximum of the pseudo-energy, and  recall 
$ k_{\a\beta}\equiv 1\star K_{\a\beta} =  \s_\a \ka_{\a\beta}$.
Finally we use again the TBA equations \eqref{tbae}  to obtain
\be\la{exE1}
\Delta E =
 \sum_{j=1}^{\Nadd} \Big(\e_{\Gadd_j} + T\log (1+{1\ov Y_\beta^{\rm max}})k_{\beta \Gadd_j}\Big) 
- \sum_{j=1}^{\Nrem} \Big(\e_{\Grem_j} + T\log (1+{1\ov Y_\beta^{\rm max}})k_{\beta \Grem_j}\Big)  \,.
\ee
Hence the dressed energy of an $\a$-string is
\be\la{eqE}
E_\a(v) = \e_\a(v) + T\log (1+{1\ov Y_\beta^{\rm max}}) k_{\beta \a}\,.
\ee 
Note that the last term in the formula is rapidity independent and does not contribute to the total energy in a particle-hole excitation. In previous studies of excitations at non-zero temperature, see e.g. \cite{KBI93}, only particle-hole excitations were considered and the dressed energies were given just by the pseudo-energies $\e_\a(v)$, and the rapidity independent term was neglected. Let us stress however that this term is important, in particular so that  in limit of infinite temperature the dressed energies  take their bare values $E_\a=\ve_\a$ as is expected. This follows as the functions $Y_\a$ become constant in the limit $T\to\infty$ because the driving terms drop out of the TBA equations \eqref{tbae}.

\subsubsection*{Momentum}

In a similar way the change in  the momentum of the excited state from the equilibrium state is given by 
\be\la{changep}
\Delta P = 
 \sum_{\a}{}\Big(\sum_{k=1}^{\tilde N_\a}\, \tilde p_{\a,k} - \sum_{k=1}^{N_\a}\, p_{\a,k}\Big) \to   p_\Tadd - p_\Trem+{1 \ov 2\pi}\zeta_\a\rho_\a\star \left|\Der{p_\a}{v}\right|\,.
\ee
Let us first remark that the momentum of a state is defined modulo $2\pi$ and thus, as the momentum of the reference state is fixed, the change in momentum is also defined modulo $2\pi$.
To simplify expression \eqref{changep} we substitute \eqref{rbr}
 into \eqref{xir} and get
 \be
{1\ov 2\pi}\zeta_\a \left|\Der{p_\a}{v}\right| + \zeta_\a K_{\a\beta}\star \rho_{\beta}=\zeta_\beta\rho_{\beta}\star K_{\beta\a} -  \p_{\a{\Tadd}}+\p_{\a{\Trem}}\,.
\ee
Multiplying by $\rho_\a$, integrating,  and taking the sum over $\a$ we find
\be
 \la{xir3}
{1 \ov 2\pi}\zeta_\a\rho_\a\star\left|\Der{p_\a}{v}\right| =-\rho_\a\star(\p_{\a{\Tadd}}-\p_{\a{\Trem}})\,,
\ee
and thus
\be\la{changep2}
\Delta P =\sum_{j=1}^{\Nadd} \big(p_{\Gadd_j} - \rho_\beta\star\p_{\beta{\Gadd_j}}\big) - \sum_{j=1}^{\Nrem} \big(p_{\Grem_j} - \rho_\beta\star\p_{\beta{\Grem_j}}\big)\,.
\ee
Hence we identify the dressed momentum of an added $\a$-string as
\be\la{eqP}
P_\a=p_\a -\rho_\beta\star \phi_{\beta\a} \,.
\ee
and a removed one with the opposite sign. 
 The dressed momentum can be used to parametrise the strings and this is discussed in appendix \ref{DrPvar}.  

To examine the range of dressed momentum it is useful to note that 
\be\la{eqdP}
\left|\Der{P_\a}{v}\right| = 2\pi (\rho_\a+\brho_\a)\,,
\ee
which is seen using eqs. \eqref{rbr} and \eqref{dlogS2}. For $\a$-strings of type 1 the range of dressed momentum is over $\big(P_\a(0) -\pi(n_\a+\bar n_\a), P_\a(0) +\pi(n_\a+\bar n_\a) \big)$ where $P_\a(0)=p_\a(0)+2 \pi n_\beta b_{\beta\a}+\pi n_\beta c_{\beta\a}$, $n_\a=1\star \rho_\a$ and $\bar n_\a=1\star \brho_\a$. On the other hand for strings of type 2 the range is split into two parts. Recall that the bare momentum of a type 2 $\a$-string has a jump at $v=0$ and  that it increases from $p_\a(\s_\a 0)$ to $p_\a(-\s_\a 0)$. The range of dressed momentum in this case is thus over  $\big( P_\a(\s_\a 0), P_\a(\s_\a 0)+\pi(n_\a+\bar n_\a) \big)$ and $\big( P_\a(-\s_\a 0) - \pi(n_\a+\bar n_\a), P_\a(-\s_\a 0) \big)$. In general this may result in a gap in the dressed momentum.

Let us remark that the dressed momenta depend on the choice of branch of $\log S_{\a\beta}$, i.e. the $b_{\a\beta}$ in eqs. \eqref{case1}, \eqref{case2}. Considering only strings of type 1 and 2 the dressed momenta \eqref{eqP} can be written as 
\be\la{eqP2}
P_\a=p_\a  -\rho_\beta\star \Theta_{\beta\a}  
	- 2\pi n_\beta b_{\beta\a}
	- \sum_{\beta \mbox{ \scriptsize  of type 1}} \pi n_\beta c_{\beta\a} \,,
\ee
where one is free to choose the integers $b_{\beta\a}$. Terms for  $c_{\beta\a}$ with $\beta$ of type 2 do not contribute as we restrict ourselves to parity invariant models and so the densities are even. It is of course possible to describe any excitation with a definite  choice of $b_{\beta\a}$, e.g. one can set all $b_{\beta\a}=0$. However in this case some excitations would have unnatural description which would require considering particle-hole excitations with zero energy contributing only to the total momentum. 
 For $n_\beta$ irrational one can achieve any value of dressed momentum by choosing  $b_{\beta\a}$ appropriately.   
  Let us remark that one is free to choose $b_{\beta\a}$ independently for each added and removed $\a$-string and one may refer to the set $b_{\beta\Gadd_j}$, $b_{\beta\Grem_j}$, where $\beta$ runs over all strings that interact with the excited string, as the branch of the excitation.

\subsubsection*{Phase shift}

Now we turn our attention to the scattering phase shift. 
Consider first the counting function for an $\a$-string of  the excited state
 \be\la{logBE3b}
L \,\s_\a \tilde z_\a(v)=\pi{\varphi_\a}+{L\, p_\a(v)}+\p_{\a\Tadd}(v)-\p_{\a\Trem}(v)+\sum_\beta\sum_{n=1}^{N_\beta}\,\p_{\a\beta}(v,v'_{\beta,n})\,,
\ee
Expanding $v'_{\beta,n}$ in the final term about its equilibrium value, replacing the sums by integrals, and noting equations (\ref{xir}, \ref{eqP}), one gets
 \be\la{shifts}
{L\, \s_\a \tilde z_\a} =\pi{\varphi_\a}+ L\,P_\a  + F_\a  \,.
\ee
 Recall that $F_\a$ here is the shift function  which is determined through the closed set of equations \eqref{eqFderv}. 
 Exponentiating equation \eqref{shifts} and evaluating it at a rapidity $v$ corresponding to a  mode number of the excited state it takes the form 
 \be\la{shifts1}
1 =e^{iL\,P_\a} e^{i( F_\a+\pi{\varphi_\a})}  \,.
\ee
 An added $\a$-string with rapidity $v$ has dressed momentum $P_\a(v)$ and so its scattering phase shift is $\delta_\a=F_\a(v)+\pi{\varphi_\a}$. Similarly a removed $\a$-string with rapidity $v$ has dressed momentum $-P_\a(v)$ and so its scattering phase shift is $\delta_\a=-F_\a(v) - \pi{\varphi_\a}$. Clearly the phase shifts $\delta$ are defined modulo $2\pi$. 

As the  equations \eqref{eqFderv} are linear it is natural to introduce the set of functions $\Phi_{\a\beta}(v,t)$ satisfying the following system of equations
\be\la{eqPhi}
\Phi_{\a\beta}=\p_{\a\beta}+{ \Phi_{\g\beta} \ov 1+Y_\g}\star K_{\g\a}\,,
\ee
where it is understood that $\Big({ \Phi_{\g\beta} \ov 1+Y_\g}\star K_{\g\a}\Big)(v,t) = \int dw  {\Phi_{\g\beta}(w,t) \ov 1+Y_\g(w)}K_{\g\a}(w,v) $.
Then $F_\a = \Phi_{\a\Tadd} - \Phi_{\a\Trem}$ and hence we refer to the $\Phi_{\a\beta}$ as dressed scattering phases. 
In terms of these functions equation \eqref{shifts1}  takes the following physically intuitive form 
\be\la{shift2}
(-1)^{\vp_\a}=e^{i P_\a(v) L}\,\prod_{j=1}^{\Nadd} e^{i\Phi_{\a\Gadd_j}(v,\tilde v_{\Gadd_j}) }\prod_{j=1}^{\Nrem} e^{-i\Phi_{\a\Grem_j}(v,\tilde v_{\Grem_j})}\,.
\ee
It is worth mentioning that the dressed scattering phases are in general not of a difference form, and in particular $\Phi_{\a\beta}(v,v)\neq 0$. This is a reflection of the fact that an excitation has nontrivial scattering with the equilibrium state.

\subsubsection*{Charge}

Each chemical potential appearing in the model is related to a conserved quantity and a corresponding charge. For a given chemical potential $\mu$ let us denote the corresponding bare charge\footnote{
For models where a magnetic field $B$ enters as a chemical potential the bare spin of an $\a$-string is generally given by $-{1\ov 2}{\partial \ve_\a\ov \partial B}$.}
 carried by an $\a$-string as $\barew_\a=-{\partial \ve_\a\ov \partial \mu}$. Furthermore let us introduce an object $\omega_\a=-{\partial \e_\a\ov \partial \mu}$, which we call the pseudo-charge of an $\a$-string. It satisfies  the following set of integral equations
\be\la{eqW}
\omega_\a = \barew_\a + {\omega_\beta\ov 1+Y_\beta}\star K_{\beta\a}\,.
\ee
The change in the total charge  of the excited state from the equilibrium state is
\be\la{changeW}
\Delta W = \sum_{\a}\Big(\sum_{k=1}^{\tilde N_\a}\, \barew_\a(\tilde v_{\a,k}) - \sum_{k=1}^{N_\a}\barew_\a(v_{\a,k})\Big) \to   \barew_\Tadd-\barew_\Trem +{\s_\a \ov 2\pi}\zeta_\a\rho_\a\star \barew_\a'\,,
\ee
where here we are being formal as $\barew_\a$ has no rapidity dependence. Indeed the final term is zero, but let us further analyse it nevertheless. 
Recalling that $ \zeta_\a\rho_{\a}=-{F_\a\ov 1+Y_\a}$, we have
\be\notag
\begin{aligned}
\zeta_\a\rho_\a\star \barew_\a' =  &  -{F_\a\ov 1+Y_\a}\star \barew_\a'=   -{F_\a\ov 1+Y_\a}\star \omega_\a' + 
	{F_\a\ov 1+Y_\a}\star K_{\a\beta} \star\big({\omega_\beta\ov 1+Y_\beta}\big)' \\
= & -{F_\a\ov 1+Y_\a}\star \omega_\a' + F_\a\star\big({\omega_\a\ov 1+Y_\a}\big)' - 	
	\phi_{\a\Tadd}\star\big({\omega_\a\ov 1+Y_\a}\big)' + \phi_{\a\Trem}\star
	\big({\omega_\a\ov 1+Y_\a}\big)'\\
= & F_\a \omega_\a\star\big({1\ov 1+Y_\a}\big)' + 2\pi\s_\a{\omega_\a\ov 1+Y_\a}\star (K_{\a\Tadd}-K_{\a\Trem}) - 2 \pi \s_\a  {\omega^{\rm max}_\a\ov 1+Y^{\rm max}_\a} (k_{\a\Tadd}-k_{\a\Trem})\,.
\end{aligned}
\ee
Here $\omega_\a^{\rm max} = \omega_\a(v^{\rm max})$ is defined similarly to  $Y_\a^{\rm max}$, both functions being evaluated at the value of $v$ corresponding to the maximum of the pseudo-energy.
This allows one to write the total change in charge as
\be\notag
\Delta W =\sum_{j=1}^{N^\Tadd} \big(\omega_{\Gadd_j}(\tilde v_{\Gadd_j}) -  {\omega^{\rm max}_\a\ov 1+Y^{\rm max}_\a} k_{\a{\Gadd_j}} \big) - \sum_{j=1}^{N^\Trem} \big(\omega_{\Grem_j}(\tilde v_{\Grem_j}) -  {\omega^{\rm max}_\a\ov 1+Y^{\rm max}_\a} k_{\a{\Grem_j}} \big) + {\s_\a \ov 2\pi}  F_\a \omega_\a\star\big({1\ov 1+Y_\a}\big)'\,.
\ee
We would like to present an interpretation of this change as
\be\la{DeltaW}
\Delta W = \sum_{j=1}^{N^\Tadd} W_{\Gadd_j}(\tilde v_{\Gadd_j}) 
	 - \sum_{j=1}^{N^\Trem} W_{\Grem_j}(\tilde v_{\Grem_j}) 
	 +\Delta W_{\rm ind}\,.
\ee
 Here the excited strings are assigned a dressed charge
\be\la{drW}
W_\a(v) = \omega_{\a}(v) -  {\omega^{\rm max}_\beta\ov 1+Y^{\rm max}_\beta} k_{\beta\a}=-{\partial E_\a\ov \partial \mu} \,,
\ee
that they carry, while the final term 
\be\la{indW}
\Delta W_{\rm ind}={\s_\a \ov 2\pi}  F_\a \omega_\a\star\big({1\ov 1+Y_\a}\big)'
\ee
is understood as an  induced charge of the system.  An interesting feature here is that the dressed charge carried by an excited string depends in general on the string's rapidity. In the limit of infinite temperature the functions $Y_\a$ become constant and the dressed charges take their bare values while the induced charge goes to zero. The zero temperature limit will be discussed in the next section where it is seen that the induced charge resides at the edge of the Fermi sea. 

The above is the interpretation we shall adopt in this paper but let us mention that the final term in eq. \eqref{DeltaW} can be redistributed among the added and removed roots using $F_\a = \Phi_{\a\Tadd} - \Phi_{\a\Trem}$. 
In particular, 
integrating by parts this final term one obtains back
\be\la{changeW2}
\Delta W =\sum_{j=1}^{\Nadd} \barew_{\Gadd_j}
	- \sum_{j=1}^{\Nrem} \barew_{\Grem_j}\,,
\ee
via the curious identity
\be
 {\omega_\beta\ov 1+Y_\beta}\star K_{\beta\a}= {\s_\a\ov2\pi}{1 \ov 1+Y_\beta}\star(\omega_\beta \Phi_{\beta\a})' \,.
\ee

One may wonder why we insist on the interpretation of eq. \eqref{DeltaW} over that of eq. \eqref{changeW2}. These are two ways of interpreting $\Delta W$ that imply different physics. That the change in charge can be split as in eq. \eqref{DeltaW} and that the dressed charge is related to the dressed energy as $W=-\Der{E}{\mu}$ is  quite convincing.  An important factor also is that spin-charge separation has been observed experimentally \cite{Ketal96} and to account for it requires an understanding of the dressing of charge that extends to non-zero temperatures. Equation \eqref{changeW2} does not provide this.

\subsection{Zero temperature}\la{zeroT}

Now we turn our attention to the limit of zero temperature. This is a special limit as  the nature of excitations may change. In section \ref{Bmodel} the zero temperature limit of the Hubbard-Shastry B-model is examined in detail and much of what is indicated here is made precise.

Examining the TBA equations \eqref{tbae} in the zero temperature limit we see that it is better to work with the pseudo-energies $\e_\a=T\log Y_\a$ rather than with the functions $Y_\a$ directly. 
Indeed, in the limit $T\to 0$ we see that the functions $Y_\a$ become singular
\be\la{T0rels}
\begin{array}{lllll}
\lim_{T\to 0}\e_\a(v) <0 &\Leftrightarrow &\lim_{T\to 0} Y_\a(v) = 0 & \Rightarrow & \lim_{T\to 0} \brho_\a(v) = 0 \,, \\
\lim_{T\to 0}\e_\a(v) >0 &\Leftrightarrow &\lim_{T\to 0} Y_\a(v) = \infty & \Rightarrow& \lim_{T\to 0} \rho_\a(v) = 0 \,.
\end{array}
\ee
Let us note that $\e_\a(v)<0$ implies that there are no holes for $\a$-strings with spectral parameter $v$ in the ground-state, whereas $\e_\a(v)>0$ implies that there are no particles of $\a$-strings with spectral parameter $v$ in the ground-state.
For each $\a$-string let us define the following subintervals of $\I_\a$
\be
\begin{aligned}
Q_\a = \{v\,:\, \e_\a(v)<0\}\,,\\
\bar Q_\a = \{v\,:\, \e_\a(v)>0\}\,.
\end{aligned}
\ee
We say that an $\a$-string is at half-filling if $Q_\a=\I_\a$, which implies from \eqref{T0rels} that there are no holes in the ground state for such strings. 
Let us next denote the boundaries between $Q_\a$ and $\bar Q_\a$. For increasing $v$ we label as $q_\a^+$ the point where $v$ goes from  $\bar Q_\a$ to  $Q_\a$, and as $q_\a^-$ the point where $v$ goes from  $ Q_\a$ to  $\bar Q_\a$. Then in the zero temperature limit
\be
{1\ov 1+Y_\a(v)} \to \begin{cases} 1 &\mbox{if } v\in Q_\a \\ 
0 & \mbox{if }  v\in \bar Q_\a\end{cases} \,, 
\ee
 and 
\be\la{YT0de}
\Big({1\ov 1+Y_\a}\Big)'(v) \to \delta(v-q_\a^+) - \delta(v-q_\a^-)
\ee
where $\delta$ is the Dirac delta function. 
The  zero temperature limit of the TBA equations \eqref{tbae} are given by
\be\la{T0bae}
\e_\a = \ve_\a + \e_\beta \star_{Q_\beta} K_{\beta\a}\,.
\ee

First consider the situation when all strings are away from half-filling. Here the problem mentioned in the paragraph above eqs. \eqref{BE3}, \eqref{BE4} does not arise and the zero temperature limit of the formalism for excitations above equilibrium is straightforward. 
The  total change in energy for an excitation, given by \eqref{exE1}, reduces to
\be\la{exE1b}
\Delta E = \sum_{j=1}^{\Nadd}\e_{\Gadd_j}(\tilde v_{\Gadd_j}) -\sum_{j=1}^{\Nrem}\e_{\Grem_j}(\tilde v_{\Grem_j}) \,,
\ee
as for each string at less than half-filling $\e_\a^{\rm max}>0$ implies $Y_\a^{\rm max}=\infty$. This is the familiar picture in which  the pseudo-energies play the role of the dressed energies.

Similarly the change in charge \eqref{DeltaW} becomes
\be\la{exC1b}
\Delta W^i = \sum_{j=1}^{\Nadd}\omega^i_{\Gadd_j}(\tilde v_{\Gadd_j}) -\sum_{j=1}^{\Nrem}\omega^i_{\Grem_j}(\tilde v_{\Grem_j})+\Delta W^i_{\rm ind} \,.
\ee
The limit of the induced charge can be taking using eq. \eqref{YT0de} giving
\be
\Delta W^i_{\rm ind}= {\s_\a\ov 2\pi}\zeta_\a(q_\a^-) \rho_\a(q_\a^-) \omega^i_\a(q_\a^-) - {\s_\a\ov 2\pi}\zeta_\a(q_\a^+) \rho_\a(q_\a^+) \omega^i_\a(q_\a^+)\,.
\ee
We thus see that the induced charge is due to the shift of the rapidities at the boundaries of the intervals $Q_\a$. This can be understood as a back-reaction of the density, which here at zero temperature occurs at the edge of the Fermi sea.

Now we turn to the situation of having some strings in the ground state at half-filling, let us say that $\e_\g^{\rm max}\leq0$ for some $\g$-strings. Here one must be careful to only consider excitations for which the $\tilde N_\g$ satisfy the selection rules \eqref{rule1} as there are no holes for $\g$-strings. Put another way, some of the $\g$-strings in the ground state may have no corresponding mode number in the excited state, due to a decrease in the range of mode numbers, and are thus necessarily removed. On the other hand an increase in the range of mode numbers will mean that there are some holes in the excited state that do not correspond to removed strings and are thus not dynamical. This situation  requires one to reconsider the nature of the excitations. 

Let us outline a convenient prescription for dealing with excitations that change the range of mode numbers of strings which are at half-filling. 
If the range increases we choose to consider only excitations for which all the extra mode numbers are filled. In our terminology this means that in such an excitation these extra mode numbers always correspond to added strings and thus all holes of  the excited state correspond to removed strings.
Obviously if the range decreases in an excitation then
 the removed mode numbers always correspond to removed strings.
We refer to such added and removed strings as singular strings. Such 
singular strings have rapidities that  approach $v^{\rm max}$ in the limit $L\to\infty$ because they correspond to mode numbers at the edges of the range. We refer to the remainder of the added and removed strings as physical strings.  Note that this prescription does not limit the freedom to capture all possible excitations. Indeed any excitation for which not all the extra mode numbers are filled can be considered as a limit of an allowed excitation where the rapidities of  the necessary number of physical removed strings approach  $v^{\rm max}$. 

An excitation could also result in an overall shift of the mode numbers. This would correspond to the removal of some singular strings at one end of the range and the addition of singular strings at the other. It can be seen however, that for each of the quantities  of interest to us,  that this transfer of singular strings is not important.

Let us thus break the added and removed strings into two types, physical and singular
\be
N_\g^{\Tadd} = N_\g^{\rm pa} + N_\g^{\rm sa} \,,\quad
N_\g^{\Trem} = N_\g^{\rm pr} + N_\g^{\rm sr}\,,
\ee
where we use $\rm p$ and $\rm s$ to denote physical and singular respectively. Let us further denote the changes in numbers of physical and singular strings as
\be
\delta N_\g^{\rm p} = N_\g^{\rm pa}- N_\g^{\rm pr}\,,\quad
\delta N_\g^{\rm s} = N_\g^{\rm sa}- N_\g^{\rm sr}\,.
\ee
Then our prescription is that
\be\la{dNs}
\delta N_\g^{\rm s}=k_{\g\beta} \delta N_\beta^{\rm p} + k_{\g\g'} \delta N_{\g'}^{\rm s}\,,
\ee
where the right hand side here is the change in the range of mode numbers of $\g$-strings found from eq. \eqref{range1}, and we use $\g'$ as a dummy index to make it clear that the sum is only over strings which are at half-filling. Let us remark that there  may be a restriction on the number of physical roots one can excite as only  solutions to eq. \eqref{dNs} for which $\delta N_\g^{\rm s}$ is an integer for each half-filled string are allowed.

Now consider again the change in energy formula  \eqref{exE1} which here takes the form
\be\la{exE1c}
\Delta E = \sum_{j=1}^{\Nadd}\big(\e_{\Gadd_j}(\tilde v_{\Gadd_j}) - \e_\g^{\rm max} k_{\g{\Gadd_j}} \big) -\sum_{j=1}^{\Nrem}\big(\e_{\Grem_j}(\tilde v_{\Grem_j})- \e_\g^{\rm max} k_{\g{\Grem_j}}\big) \,.
\ee
Splitting the strings between their physical and singular subsets this becomes
\be
\begin{aligned}
\Delta E =& \sum_{j=1}^{N^{\rm pa}}\e_{\Gadd_j}(\tilde v_{\Gadd_j}) 
-\sum_{j=1}^{N^{\rm pr}} \e_{\Grem_j}(\tilde v_{\Grem_j})
 -\e_\g^{\rm max} k_{\g\a} \delta N^{\rm p}_\a
  + \e_\g^{\rm max} \delta N^{\rm s}_\g
-\e_\g^{\rm max} k_{\g\g'} \delta N^{\rm s}_{\g'} \\
=&\sum_{j=1}^{N^{\rm pa}}\e_{\Gadd_j}(\tilde v_{\Gadd_j}) 
-\sum_{j=1}^{N^{\rm pr}} \e_{\Grem_j}(\tilde v_{\Grem_j})\,,
\end{aligned}
\ee
where all the constant terms have cancelled due to \eqref{dNs}.
The singular strings may also have non-zero dressed momentum and non-trivial dressed scattering. These can be redistributed  among the physical strings according to the solution of \eqref{dNs}.
For example, if the solution to eq. \eqref{dNs} is $\delta N_\g^{\rm s} =  f_{\g\a}\delta N_\a^{\rm p}$ then the dressed momentum and dressed scattering take the following form for the half-filled phase
\be\la{redist}
P_\a^{\rm h.f.} = P_\a + P_\g(v^{\rm max})f_{\g\a}\,,\quad
\Phi_{\a\beta}^{\rm h.f.}(v,t) = \Phi_{\a\beta}(v,t) + \Phi_{\a\g}(v,v^{\rm max})f_{\g\beta}\,.
\ee
Finally let us consider again the change in charge. 
As for the energy, the contributions of the singular roots cancel all constant terms appearing in the dressed charge \eqref{drW}. 
Also the derivative in \eqref{indW} is zero at zero temperature for half-filled strings and so such strings do not give rise to an induced charge of the system. The formula for the change in charge thus takes the form
 \be\notag
\Delta W =\sum_{j=1}^{N^{\rm pa}} \omega_{\Gadd_j}(\tilde v_{\Gadd_j}) - \sum_{j=1}^{N^{\rm pr}} \omega_{\Grem_j}(\tilde v_{\Grem_j}) +\Delta W_{\rm ind}^{\rm a.h.f} \,,
\ee
where $\Delta W_{\rm ind}^{\rm a.h.f}$ denotes the induced charge due to the back-reaction of the strings which are away from half-filling.

\subsection{Summary of the excited state formalism}
\la{summary}

Let us summarise the main features of the excited state formalism developed in this section.
In the thermodynamic limit the Bethe equations become the equations for densities \eqref{rbr}. Requiring the free energy to be minimised yields the TBA equations \eqref{tbae}. This closed set of non-linear integral equations on $\brho_\a\ov\rho_\a$ allows one to determine the equilibrium densities. Excitations of a finite number of strings above the equilibrium state can be investigated by examining the shifts of the roots \eqref{defxi} arising due to an excitation. These shifts also satisfy a closed set of non-linear integral equations \eqref{xir}, and with the aid of these equations all the features of an excitation can be extracted.

Let us clear the notations of the previous subsections and parametrise an excited state  by $N^\Radd$ added particles with rapidities $v_{\Radd_k}$, $k=1,\ldots,N^\Radd$,  and  by $N^\Rrem$ holes with rapidities $v_{\Rrem_k}$, $k=1,\ldots,N^\Rrem$.
Here the indices ${\Radd_k}$ and ${\Rrem_k}$ include the information of the type of string. The excitation can be encoded in a set of Bethe equations for which the pseudo-vacuum is the equilibrium state
\begin{align}\la{shift3}
(-1)^{\varphi_a}=& e^{i P_{\Radd_k}(v_{\Radd_k}) L}\,\prod_{j=1 }^{N^\Radd} e^{i\Phi_{\Radd_k \Radd_j}(v_{\Radd_k}, v_{\Radd_j}) }\prod_{j=1}^{N^\Rrem} e^{-i\Phi_{\Radd_k \Rrem_j}(v_{\Radd_k}, v_{\Rrem_j})}\,,\\
(-1)^{-\varphi_a}=& e^{i P_{\Rrem_k}(v_{\Rrem_k}) L}\,\prod_{j=1 }^{N^\Rrem} e^{i\Phi_{\Rrem_k \Rrem_j}(v_{\Rrem_k}, v_{\Rrem_j}) }\prod_{j=1}^{N^\Radd} e^{-i\Phi_{\Rrem_k \Radd_j}(v_{\Rrem_k}, v_{\Radd_j})}\,.
\end{align}
Let us stress that these Bethe equations are only valid for large $L$. The momentum is given through eq. \eqref{eqP}
\begin{align}
 {P_\a}&=p_\a - \rho_\beta\star \p_{\beta\a}\quad\mbox{for particle excitations,} \\
  {P_\a}&=-p_\a + \rho_\beta\star \p_{\beta\a}\,\quad\mbox{for hole excitations,} 
 \end{align}
 and the scattering phases $\Phi_{\a\beta}$ are determined through the closed set of equations \eqref{eqPhi}.
Due to interactions with the equilibrium state the energy of each excited root gets dressed  \eqref{eqE}
\begin{align}
 {E_\a}&=\e_\a + T\log (1+{1\ov Y_\beta^{\rm max}}) k_{\beta\a} \quad\mbox{for particle excitations,} \\
  {E_\a}&=- \e_\a - T\log (1+{1\ov Y_\beta^{\rm max}}) k_{\beta\a}\quad\mbox{for hole excitations.} 
 \end{align}
 Here $\e_\a=T\log Y_\a$ are the psuedo-energies and the constants $Y_\a^{\rm max}$ and $k_{\a\beta}$ are defined after eq. \eqref{changeE3}.
Similarly, for a conserved quantity the corresponding charge of each excited root gets dressed and is given by \eqref{drW}
\begin{align}
 {W_\a}&=\omega_\a - {\omega_\beta^{\rm max}\ov 1+Y_\beta^{\rm max}}
 k_{\beta\a}= -\Der{E_\a}{\mu} \quad\mbox{for particle excitations,} \\
  {W_\a}&=- \omega_\a +  {\omega_\beta^{\rm max}\ov1+ Y_\beta^{\rm max}} k_{\beta\a}= -\Der{E_\a}{\mu} \quad\mbox{for hole excitations,} 
 \end{align}
where $\omega_\a$ are the pseudo-charges determined through the closed set of equations \eqref{eqW} and the constant $\omega_\a^{\rm max}$ is defined above eq. \eqref{DeltaW}.
 The total change in charge for the excitation also has a contribution $\Delta W_{\rm ind}$ given in eq.  \eqref{indW}. This is an induced charge of the system that is not carried by the excited roots, but  rather is due to a back-reaction of the densities. 

 In the limit of infinite temperature $T\to\infty$ the dressed energy and dressed charge take their bare values as one expects. 
 The limit of zero temperature $T\to0$ requires special attention and is discussed in detail in section \ref{zeroT}. 
 If some strings are at half-filling there may be restrictions on allowed excitations  given through eq. \eqref{dNs}  and their nature may be altered.  Regardless of whether this is necessary the dressed energy and dressed charge of excited roots take their pseudo values in the zero temperature limit.



\section{Hubbard-Shastry models}\la{HSmods}

Now we wish to examine the excitations of the Hubbard-Shastry models. An introduction to these models can be found in \cite{FQ11}, as well as an overview of their equilibrium state. We will focus here on the A- and B-models as the excitations of the Hubbard model 
have been investigated elsewhere,  see for example \cite{bookH} and references therein.
Not alone are these models of great interest but also the B-model in particular provides a good illustrative  example for the application of the formalism developed in section \ref{formalism}. We will restrict our attention to excitations at zero temperature.

 We begin with a brief review of the Hubbard-Shastry A- and B-models. Let us first write the following general Hamiltonian for a one-dimensional lattice of length $L$
 \begin{align}\notag
  {\mathbf H} &= \sum_{j=1}^L\, \Big( {\mathbf T}_{j,j+1}+\ka_{\rm H}\, {\mathbf V}^{\rm H}_{j,j+1}
  		+ \ka_{\rm CC}\, {\mathbf V}_{j,j+1}^{\rm CC} + \ka_{\rm SS}\,{\mathbf V}_{j,j+1}^{\rm SS} + \ka_{\rm PH}\, {\mathbf V}_{j,j+1}^{\rm PH}  \Big) -\mu {\mathbf N}- 2B {\mathbf S}^z\,,\\ \la{genHam}
  {\mathbf T}_{j,k} &=  - \sum_{\sigma} \Big[ \cd_{j,\sigma} \cm_{k,\sigma}
  \big( \tau_{0} +\tau_{1} \n_{j,-\sigma} +\tau_{2} \n_{k,-\sigma} + \tau_{3}\n_{j,-\sigma} \n_{k,-\sigma}\big) \\
 & \qquad\qquad  + \cd_{k,\sigma} \cm_{j,\sigma}
  \big( \bar\tau_{0} +\bar\tau_{1} \n_{j,-\sigma} +\bar\tau_{2} \n_{k,-\sigma} + \bar\tau_{3}\n_{j,-\sigma} \n_{k,-\sigma}\big) \Big]\,.\notag
 \end{align}
 Here the canonically anticommuting fermionic operators $\cd_{j,\sigma}$  create and $\cm_{j,\sigma}$  annihilate electrons of spin $\sigma = \uparrow$ or $\sigma =\downarrow$ at the $j$-th site of the lattice. The operator $\n_{j,\sigma} =\cd_{j,\sigma}\cm_{j,\sigma}$ is the local particle number operator for electrons of spin $\s$ at site $j$, $\mu$ is the chemical potential and ${\mathbf N}=\sum_{j=1}^{L} \n_{j,\uparrow} + \n_{j,\downarrow}$, and $B$ is a magnetic field coupling to the $z$-component of the spin operator $ {\mathbf S}^z = {1\ov 2} \sum_{j=1}^{L} \n_{j,\uparrow} - \n_{j,\downarrow}$.
 The respective Hubbard, charge-charge, spin-spin and pair hopping interactions are
 \be\notag
 \begin{aligned}
  {\mathbf V}^{\rm H}_{j,k} &={1\ov 2} \,\big(\n_{j,\uparrow}-{1\ov 2}\big)\big(  \n_{j,\downarrow}-{1\ov 2}\big) +{1\ov 2} \,\big(\n_{k,\uparrow}-{1\ov 2}\big)\big(  \n_{k,\downarrow}-{1\ov 2}\big)-{1\ov 4}\,,\\
 {\mathbf V}_{j,k}^{\rm CC} & = \,{\boldsymbol \eta}_j^z\, {\boldsymbol \eta}_k^z-\frac{1}{4}\ =\ \frac{1}{4}\left(\n_{j,\uparrow} + \n_{j,\downarrow}-1)(\n_{k,\uparrow} + \n_{k,\downarrow}-1 \right)-\frac{1}{4} \,,\\
 {\mathbf V}_{j,k}^{\rm SS} &= \frac{1}{2}(	{\mathbf  S}_j^+\, {\mathbf  S}_k^-+{\mathbf  S}_j^- \,{\mathbf  S}_k^+ )
 	+ {\mathbf  S}_j^z\, {\mathbf  S}_k^z\\
& =  \frac{1}{2}(\cd_{j,\uparrow} \cm_{j,\downarrow} \cd_{k,\downarrow} \cm_{k,\uparrow}+\cd_{j,\downarrow} \cm_{j,\uparrow} \cd_{k,\uparrow} \cm_{k,\downarrow} )
 	+\frac{1}{4} \left(\n_{j,\uparrow} - \n_{j,\downarrow} \right)\left(\n_{k,\uparrow} - \n_{k,\downarrow} \right)\,,\\
  {\mathbf V}_{j,k}^{\rm PH} & = \frac{1}{2}( {\boldsymbol \eta}_j^+\, {\boldsymbol \eta}_k^- +{\boldsymbol \eta}_j^-\, {\boldsymbol \eta}_k^+ )\ =\  \frac{1}{2}(\cd_{j,\uparrow} \cd_{j,\downarrow} \cm_{k,\downarrow} \cm_{k,\uparrow} +\cd_{k,\uparrow} \cd_{k,\downarrow}   \cm_{j,\downarrow}\cm_{j,\uparrow})\,.
\end{aligned}
\ee 
The A-model and B-models each have one free coupling constant $\nu$ that is related to the parameters given in \eqref{genHam} as 
\bean
&\ka_{\rm H} = {2\cosh2\nu\ov\cosh\nu}\,,\quad\ka_{\rm CC} =-\ka_{\rm SS} =\ka_{\rm PH} = {2\ov\cosh\nu}\,,\\
&\tau_{0}=1\,,\quad  \tau_{1}= \bar\tau_{2} =-1-i\tanh\nu\,,\quad  \tau_{3}=2i\tanh\nu\,,
\eean
for the A-model, and 
\bean
&\ka_{\rm H} = -\ka_{\rm CC} =\ka_{\rm SS} =\ka_{\rm PH} = 2\tanh\nu\,,\\
&\tau_{0}=1\,,\quad \tau_{1}=\tau_{2}=-1+\mbox{sech}\,\nu \,,\quad  \tau_{3}=-2\tau_{1}\,.
\eean
for the B-model.

Now we present the models in their diagonalised form which is how we will view them in this paper. Here it is more convenient to reparametrise the coupling constant as 
\be \notag
\uh=\sinh\nu\,.
\ee
Their Bethe equations are \cite{MaRa}
\be\la{BEHS}
\begin{aligned}
1 =& e^{iL\, p(v_k)}
\prod_{j=1}^{M}\frac{v_k-w_j-i\,\uh}{v_k-w_j+i\,\uh}\,, & k=1,\ldots,N \, \le L\,,  \\
-1=&\prod_{j=1}^{N}\frac{w_k-v_j-i\,\uh}{w_k-v_j+i\,\uh}
 \prod_{l=1}^{M}\frac{w_k-w_l+2i\,\uh}{w_k-w_l-2i\,\uh}\,,& k=1,\ldots,M  \le {N\ov2}\,,
\end{aligned}
\ee
where $e^{i p(v)}$ for each of the models (and for completeness for the Hubbard model also) are given in Table \ref{HSpE}, along with the dispersion relations $\E(p)$. 
Note that rapidity variable $v$ of momentum carrying roots is related to the momentum through $y$ with $v={1\ov 2}(y+1/y)$, and so $e^{i p(v)}$ is a double valued function of $v$. We refer to these roots as $y$-particles and they are discussed in detail in appendix \ref{HSy}. 
\begin{table}[htdp]
\begin{center}
\begin{tabular}{c|cccc}
  & A-model  & B-model  &  Hubbard\\
\hline
 $e^{ip(v)}$   &  $i{ 1+ y x^+ \ov y - x^+}$   & ${y +x^+ \ov y -x^+}$& ${ i\, y}$\\
 $\E(p)$  & $-2 \cos p - 2\sqrt{1+\uh^2}$ & $-2 \cos p$ & $-2 \cos p - 2\uh$
\end{tabular}
\end{center}
\vspace{-0.5cm}
\caption{\small The momenta and dispersion relations for the Hubbard-Shastry models. \newline Here $v={1\ov 2}(y+1/y)$ and $x^+ = i(\uh+\sqrt{1+\uh^2})$.}
\label{HSpE}
\end{table}

The string hypothesis for the behaviour of the Bethe roots in the thermodynamic limit is that each root is a member of one of the following types of strings \cite{Takahashi,Takbook}
\bei
\item $y$-particle: a charge 1, spin-up momentum carrying particle,
\item $M|{vw}$-string: a charge $2M$, zero spin momentum carrying bound state, 
\item $M|{w}$-string: a zero charge, spin $-M$ bound state,
\eei
where $M$ denotes a positive integer. 
The momentum and the dispersion relation for the $M|{vw}$-strings are 
\be\notag
\begin{aligned}
p_{M|vw}(v)&=\sum_{j=1}^{M} p_+\big(v+(2j-M)i\uh\big)+p_-\big(v-(2j-M)i\uh\big)\,,\\
 \E_{M|vw}(v)&=\sum_{j=1}^{M} \E_+\big(v+(2j-M)i\uh\big)+\E_-\big(v-(2j-M)i\uh\big)\,,
\end{aligned}
\ee
and  $\s_{1|vw}=\mbox{sign}\big(\Der{p_{M|vw}}{v}\big)=-1$.
Here we have adopted the clean notations of \cite{AFS09}, which are compared with the more conventional notations, of say \cite{bookH}, in appendix \ref{conventions}.

\begin{figure}[htbp]
\begin{center}
\includegraphics[width=0.83\textwidth]{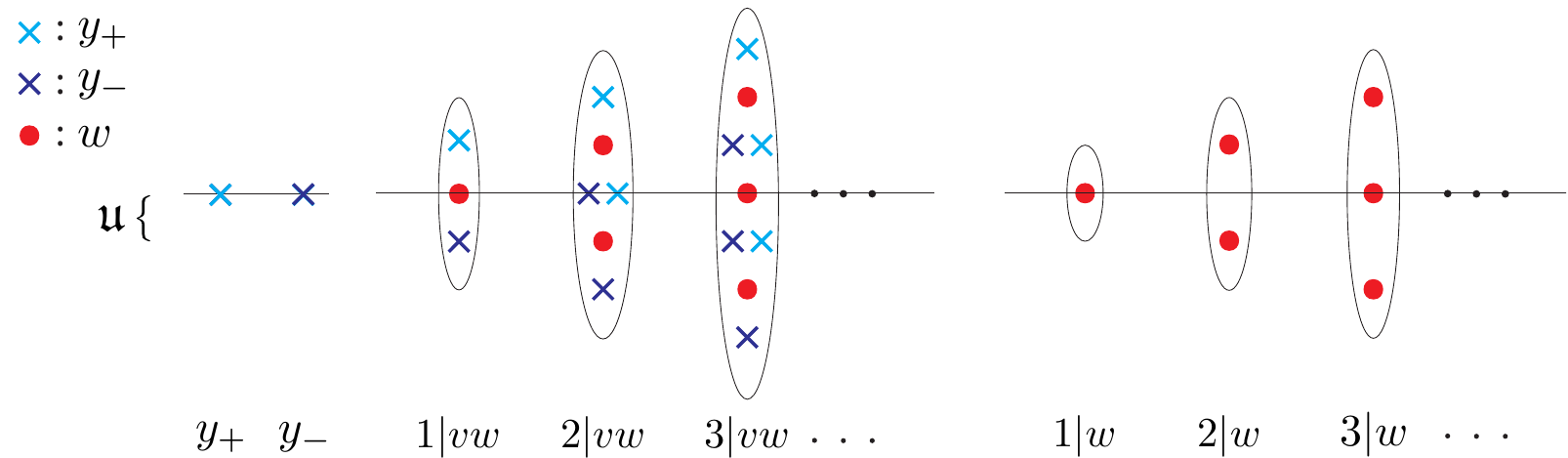}
\caption{\small (Colour online) An illustration of string hypothesis for the Hubbard-Shastry models. The horizontal line represents the real axis while the imaginary direction extends vertically. The $y_\pm$ refer to the two branches of the $y$-particle and the $\times$ mark the corresponding value of the $v$-rapidity variable, which take values on $(-1,1)$ for the A-model, and on $(-\infty,-1)\cup(1,\infty)$ for the B-model. The circles mark the rapidities of $w$ roots. The $M|vw$ and $M|w$ strings have real centres that take values on $\Reals$. The $M|vw$-strings are momentum carrying and their momentum $p_{M|vw}$ and dispersion $\E_{M|vw}$ are obtained by summing the contributions of the roots of which they are composed. 
}
\label{stringhyp}
\end{center}
\end{figure}

These strings constitute the particle content of the model in the thermodynamic limit and they are summarised in figure \ref{stringhyp}.
Re-writing the Bethe equations in terms of the string solutions they take the form\footnote{These correct a minor error in eqns. (A.11-A.13) of \cite{FQ11}.}
\be\la{HSsBE}
\begin{aligned}
1= & \ e^{iL\, p_y(v_{y,k})} 
 \prod_{M=1}^{\infty} \prod_{j=1}^{N_{M|vw}}S_M(v_{y,k}-v_{M,j}) \,
  \prod_{N=1}^{\infty} \prod_{l=1}^{N_{N|w}}S_M(v_{y,k}-w_{N,l}) \,,\\
-1 = & \ e^{iL\, p_{M|vw}(v_{M,k}) }
\prod_{j=1}^{N_{y}}S_M(v_{M,k} -v_{y,j})
 \prod_{N=1}^{\infty} \prod_{l=1}^{N_{N|vw}}S_{MN}(v_{M,k}-v_{N,l})\,,\\
 -1 =&\ \prod_{j=1}^{N_{y}}S_M(w_{M,k} -v_{y,j}) \prod_ {N=1}^{\infty} \prod_{l=1}^{N_{N|w}}{1\ov S_{MN}(w_{M,k}-w_{N,l})}\,.
 \end{aligned}
 \ee
 The S-matrices are given in appendix \ref{conventions}. 
These are the string Bethe equations which represent the starting point, eq. \ref{BE1}, of the formalism developed in section \ref{formalism}. 

There are two conserved quantities corresponding to charge and spin, and their respective chemical potentials are $\mu$ and $B$. The energies carried by the strings are  influenced by these chemical potentials and are
\be
\ve_y=\E_y-\mu-B\,,\quad \ve_{M|vw} = \E_{M|vw} -2M\mu\,,\quad \ve_{M|w}=2M B\,.
\ee
We denote the bare charge of an $\a$-string by $\barew^{\rm c}_\a=-{\partial \ve_\a\ov \partial \mu}$ and the bare spin by $\barew^{\rm s}_\a=-{1\ov 2}{\partial \ve_\a\ov \partial B}$.

\subsection{A-model}
Now we examine the zero temperature excitations of the A-model. Let us first  recall the ground state. 
This is determined by the zero temperature limit of the TBA equations \eqref{tbae}
\be\la{AT0TBA}
\begin{aligned}
\e_y&=\E_y-\mu-B\,,\\
\e_{M|vw}&=\E_{M|vw}-2M \mu + \e_y \circledast_{Q_y}\, K_M \,,\\
\e_{M|w}&=2MB + \e_y \circledast_{Q_y}\, K_M \,.
\end{aligned}
\ee
The dispersions $\E_y$ and $\E_{1|vw}$ are plotted as functions of $v$ in Figure \ref{EyvwA}.
\begin{figure}[t]
\includegraphics[width=0.48\linewidth]{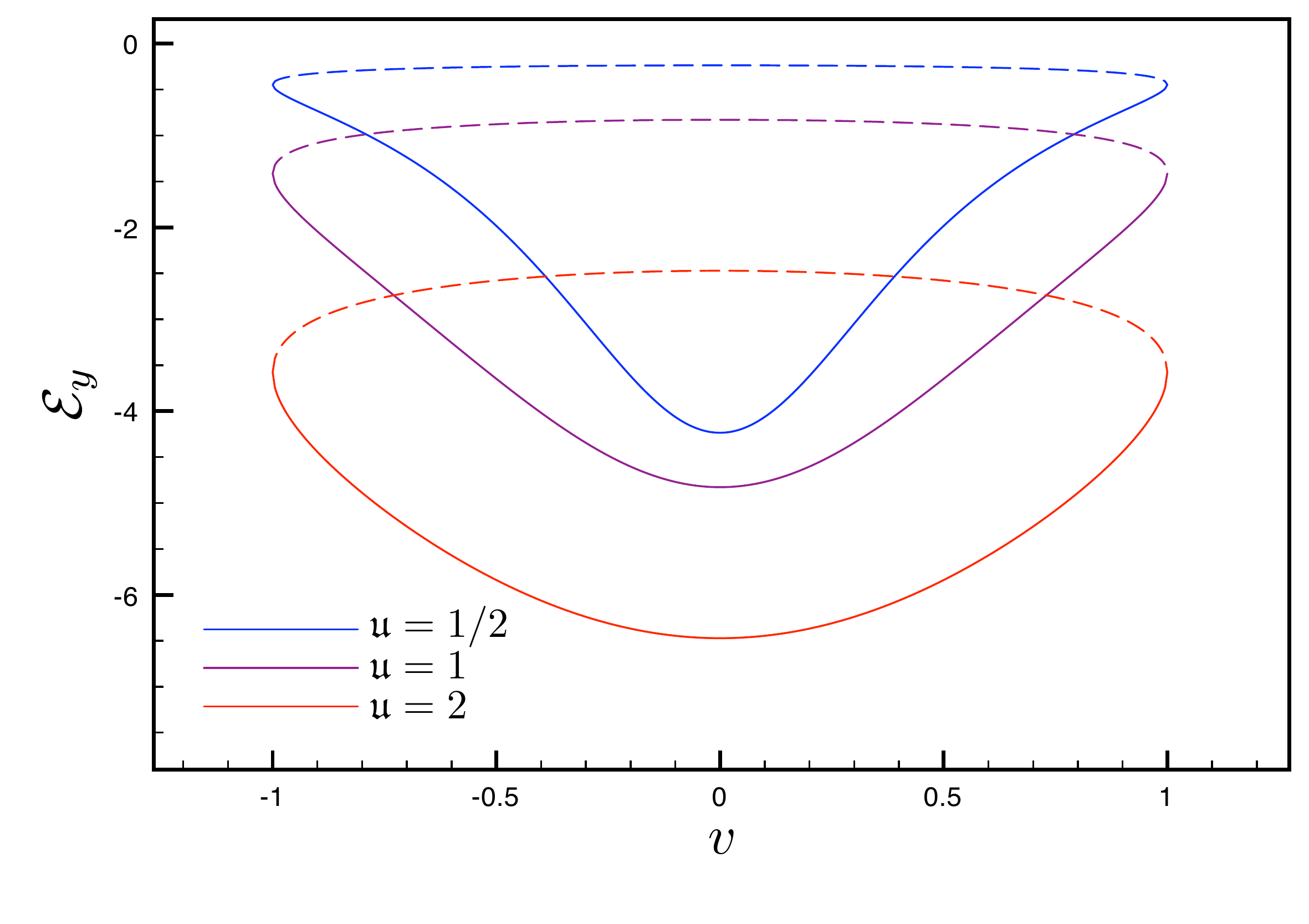}
\hfill
\includegraphics[width=0.48\linewidth]{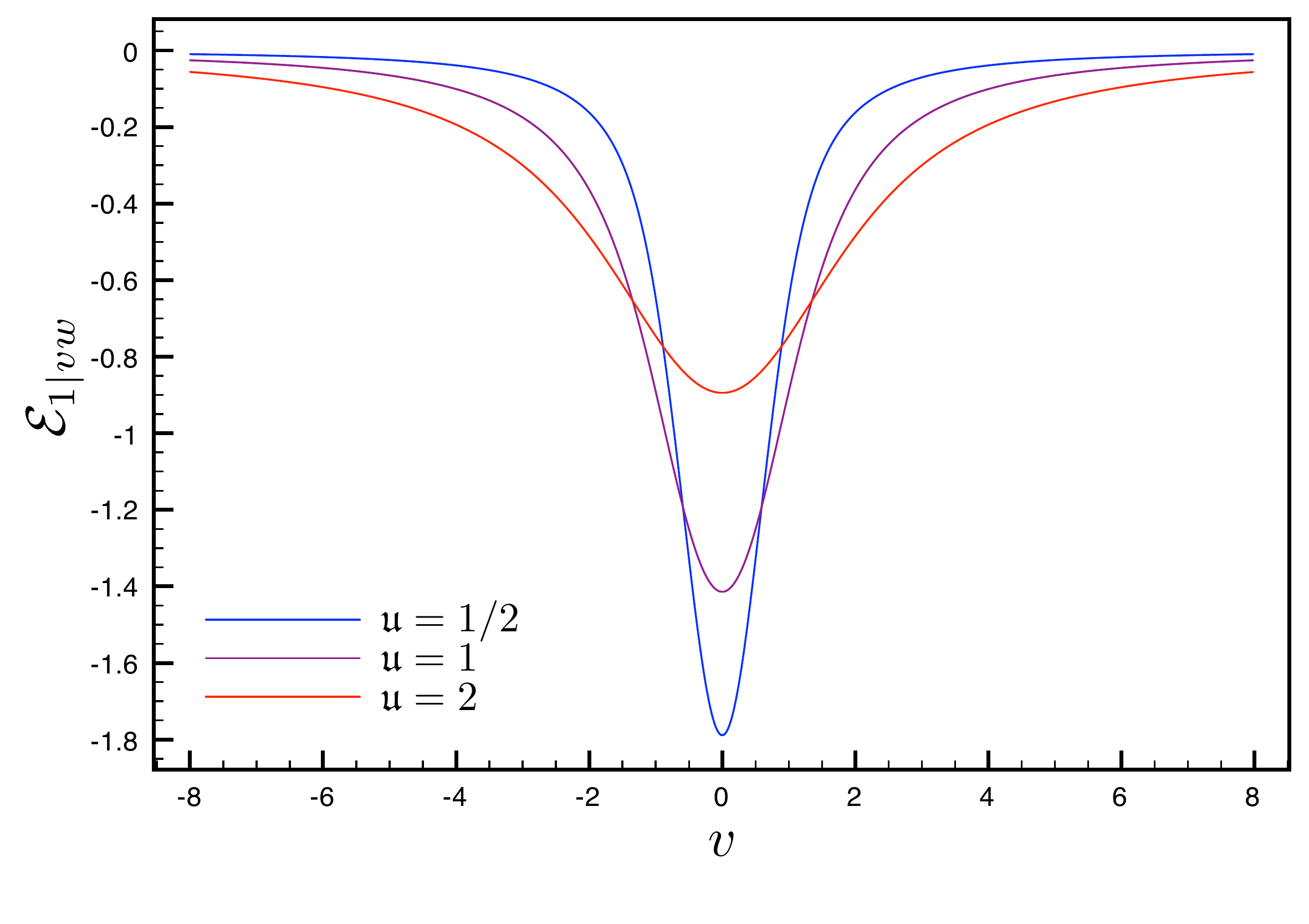}
\vspace{-0.5cm}
\caption{{\small  A-model: Plots of $\E_y(v)$ and $\E_{1|vw}(v)$ for $\uh=1/2,1,2$. In the plots of $\E_y(v)$ the $y_+$ branch is represented by a solid line and the $y_-$ branch is represented by a dashed line. }}
  \la{EyvwA}
\end{figure}
The $y$-particles are the only strings which have a non-zero density and so the ground state is spin polarised.  As $y$-particles have charge 1 and spin ${1\ov 2}$ the magnetisation of the ground state is equal to half of the density.
 The phase diagram\footnote{We restrict our attention to the quadrant $\mu\leq0$ and $B\geq0$ as the other quadrants are related by symmetries.} is presented in Figure \ref{phasediagA}.
\begin{SCfigure}[2.6][t]
  \centering
  \includegraphics[width=0.21\textwidth]%
    {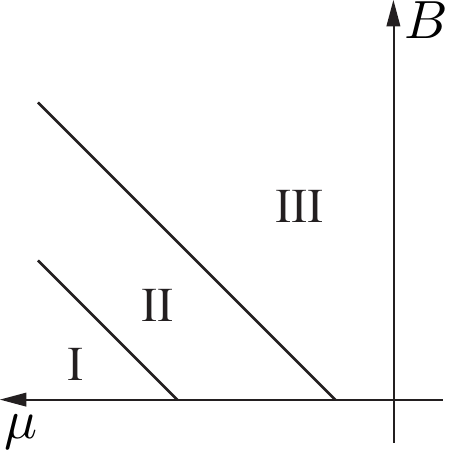}
    \hspace{1em}
  \caption{\small A-model: Zero temperature phase diagram  in the $\mu B$-plane. The phases identified are: I) empty band, II) partially filled and spin polarised band, III) half-filled and spin-polarised band. The line separating phases I and II is $\mu+B=2-2\sqrt{1+\uh^2}$, and the line separating phases II and III is $\mu+B=-2-2\sqrt{1+\uh^2}$. }
  \la{phasediagA}
\end{SCfigure}
The ground state is empty in phase I as the energy cost of having electrons in the ground state is too high. In phase II $y$-particles enter the ground state and the filling increases with increasing $\mu$, up to the boundary with phase III where the ground state becomes half-filled.
Analysing the TBA equations \eqref{AT0TBA} one can see that the $M|vw$-strings are type 1 strings while 
the $M|w$-strings are of type 2. 
The $y$-particles should be treated as type 1 strings as discussed in appendix \ref{HSy} and so the phase shifts are
\be\notag
\phi_{y,M|vw}(v,t)=\pi b_{y,M|vw} +\Theta_M(v-t)\,,\quad
\phi_{y,M|w}(v,t)=\pi b_{y,M|w} +\Theta_M(v-t)\,,
\ee
and  the dressed momenta are
\be\la{DrPA}
\begin{aligned}
P_y=p_y\,,\quad& P_{M|vw}=p_{M|vw} - \Der{p_y}{v}\circledast_{Q_y}\Theta_M - \pi n_y (2 b_{y,M|vw} +1) \,,\\
&P_{M|w}= - \Der{p_y}{v}\circledast_{Q_y}\Theta_M - \pi n_y (2 b_{y,M|w}+1)\,,
\end{aligned}
\ee
where $b_{y,M|w}$ and  $b_{y,M|vw}$ are the integers that determine the branch of the excitation.

Let us first consider excitations above the half-filled phase III. 
One does not have to introduce singular strings through eq. \eqref{dNs} as  the range of mode numbers for $y$-particles $N_y+\bar N_y=L$ does not change for an excitation. That $k_{y\beta}=0$ for all $\beta$-strings is a reflection of this. Moreover the dressed energies can be written explicitly as
\be\la{hfAT0TBA}
\e_y=\E_y-\mu-B\,,\quad
\e_{M|vw}=-2M \mu \,,\quad
\e_{M|w}=-\E_{M|vw}+2MB\,,
\ee
while the dressed momenta are branch independent and simplify to
\be
P_y=p_y\,,\quad P_{M|vw}=0\,,\quad P_{M|w}=-p_{M|vw}\,.
\ee
Here the identities \eqref{iddpE} have been used.
The dressed charge and spin are equal to their bare values. 
Thus the $y$-particles behave as electrons and are gapped, and so there is an energy cost to remove them from the ground state. 
Their dressed energy and momentum take their bare values and so are related as 
\be
\e_y = - 2 \cos P_y - 2\sqrt{1+\uh^2}-\mu-B\,.
\ee
The $1|w$-string is the magnon and the $M|w$-strings are their bound states. In $B=0$ magnetic field they have quadratic dispersion at low energies, $\e_{M|w}\sim {\sqrt{1+\uh^2}-\uh \ov M}P_{M|w}^2$, 
 while in a $B>0$ magnetic field they are gapped. The $M|vw$-strings are not dynamical. In the strong coupling $\uh\to\infty$ limit the energy gap for removing an electron goes to infinity and the physics becomes that of the ferromagnetic spin chain\footnote{
The ferromagnetic spin-chain Hamiltonian appears at order $1\ov\uh$ and so it is necessary to rescale the energies appropiately.} \cite{FQ11}.

Now we consider excitations above phase II where the filling ranges between 0 and 1. 
The dressed energies are given by \eqref{AT0TBA} and the dressed momentum by \eqref{DrPA}, and here the choice of branch becomes important.  For convenience we choose $b_{y,M|vw}=0$, $b_{y,M|w}=-1$ for all $M$. The range of dressed momentum is then $(-\pi,\pi)$ for $y$-particles, $(0,2\pi- 2\pi n_y)$ for $M|vw$-strings and $(0,2\pi n_y)$ for $M|w$-strings. These should be considered modulo $2\pi$ and it is convenient to plot the dressed momentum of the $M|vw$-strings and  the $M|w$-strings in the range $(0,2\pi)$. It should be kept in mind that the other branches of the excitations are obtained by shifts of $2\pi n_y$. Plots of dressed energy as functions of dressed momentum for $y$-particles, $M|w$-strings and $M|vw$-strings  for various filling at $B=0$ and $\uh=1$ are given in  Figures  \ref{EPyvwA} and \ref{EPDscwA}.
\begin{figure}[t]
\includegraphics[width=0.50\linewidth]{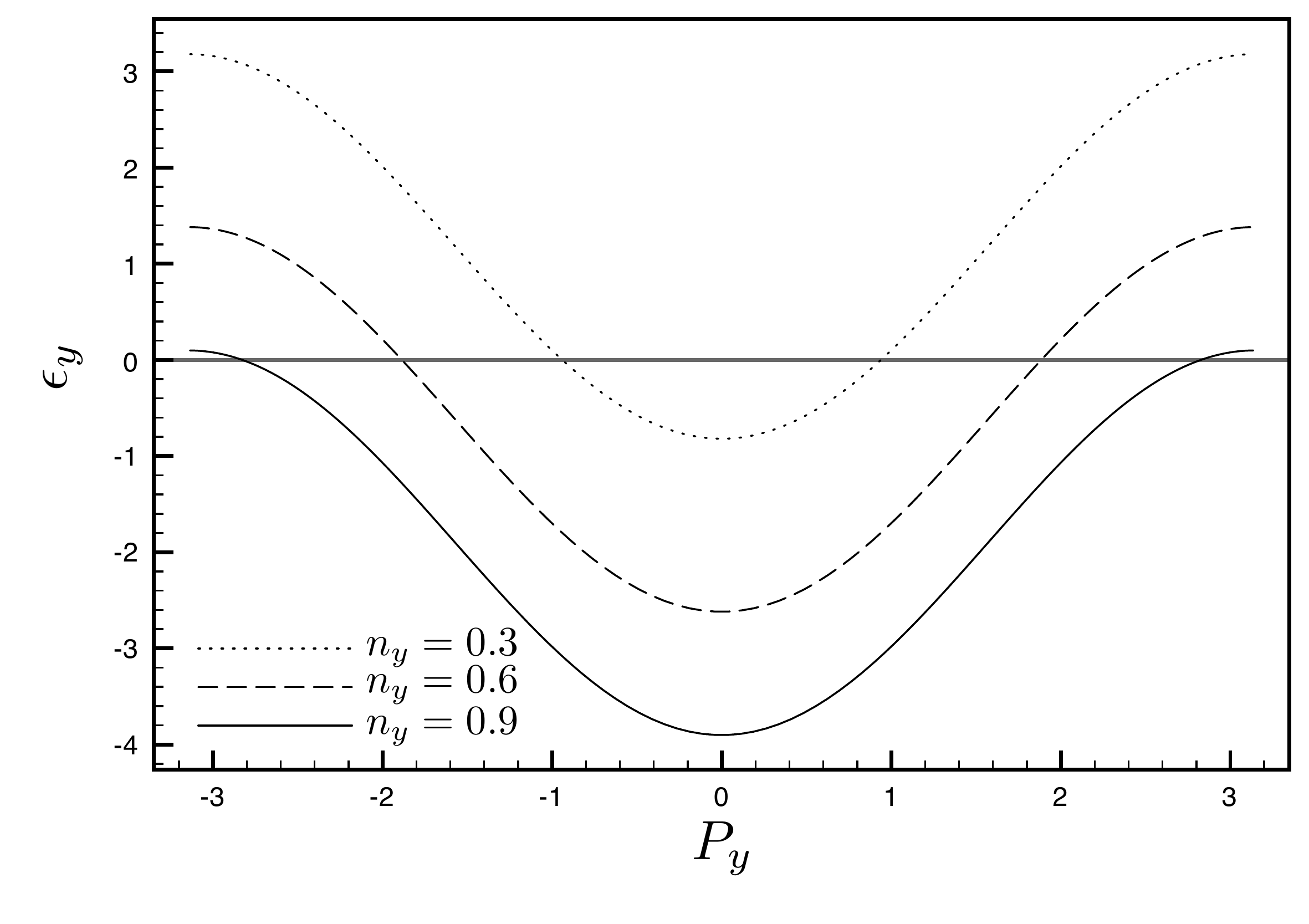}
\hfill
\includegraphics[width=0.50\linewidth]{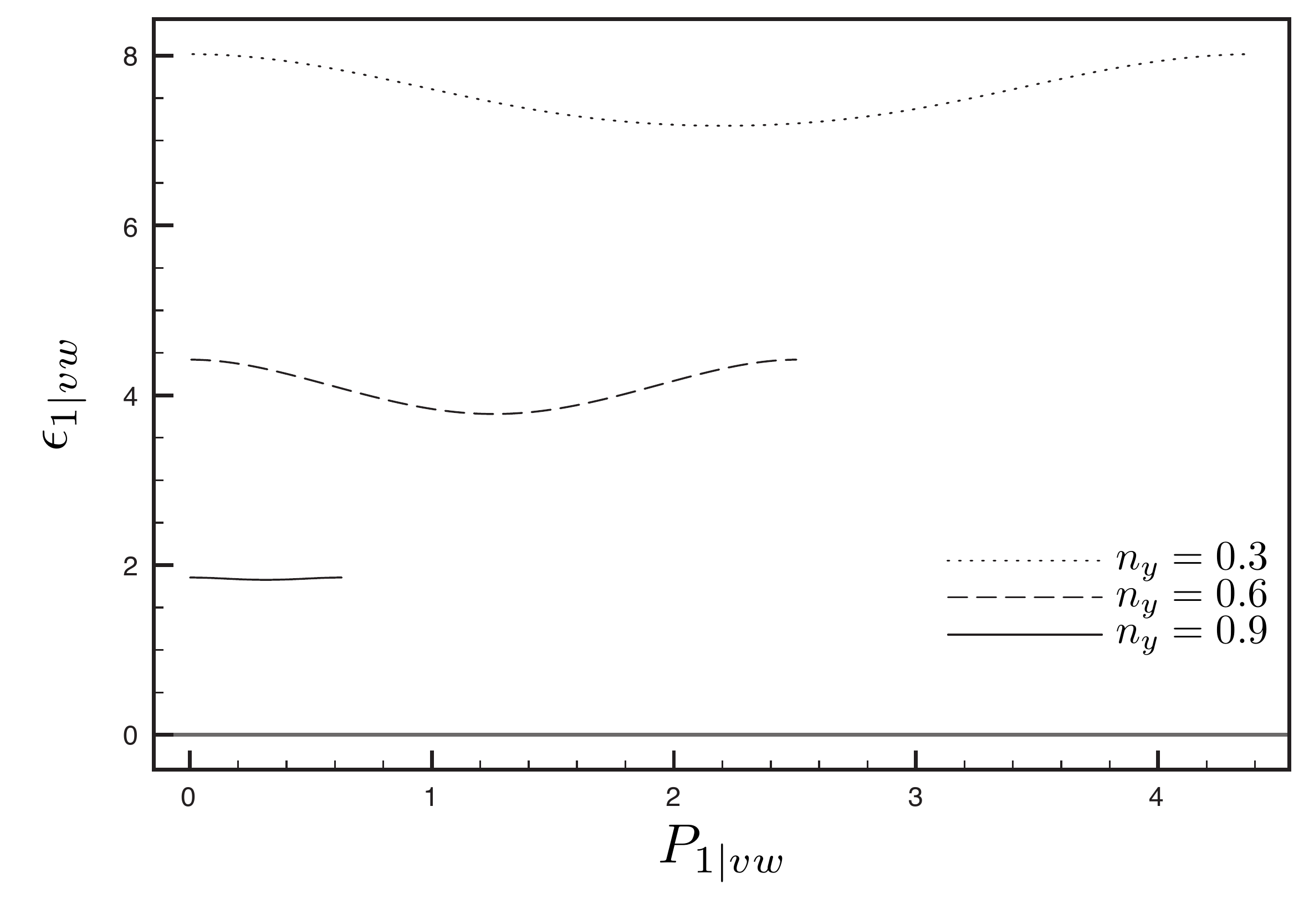}
\vspace{-0.5cm}
\caption{{\small  A-model: Plots of $\e_y(P_y)$ and $\e_{1|vw}(P_{1|vw})$  at $\uh=1$ and $B=0$. }}
  \la{EPyvwA}
\end{figure}
\begin{figure}[t]
\includegraphics[width=0.48\linewidth]{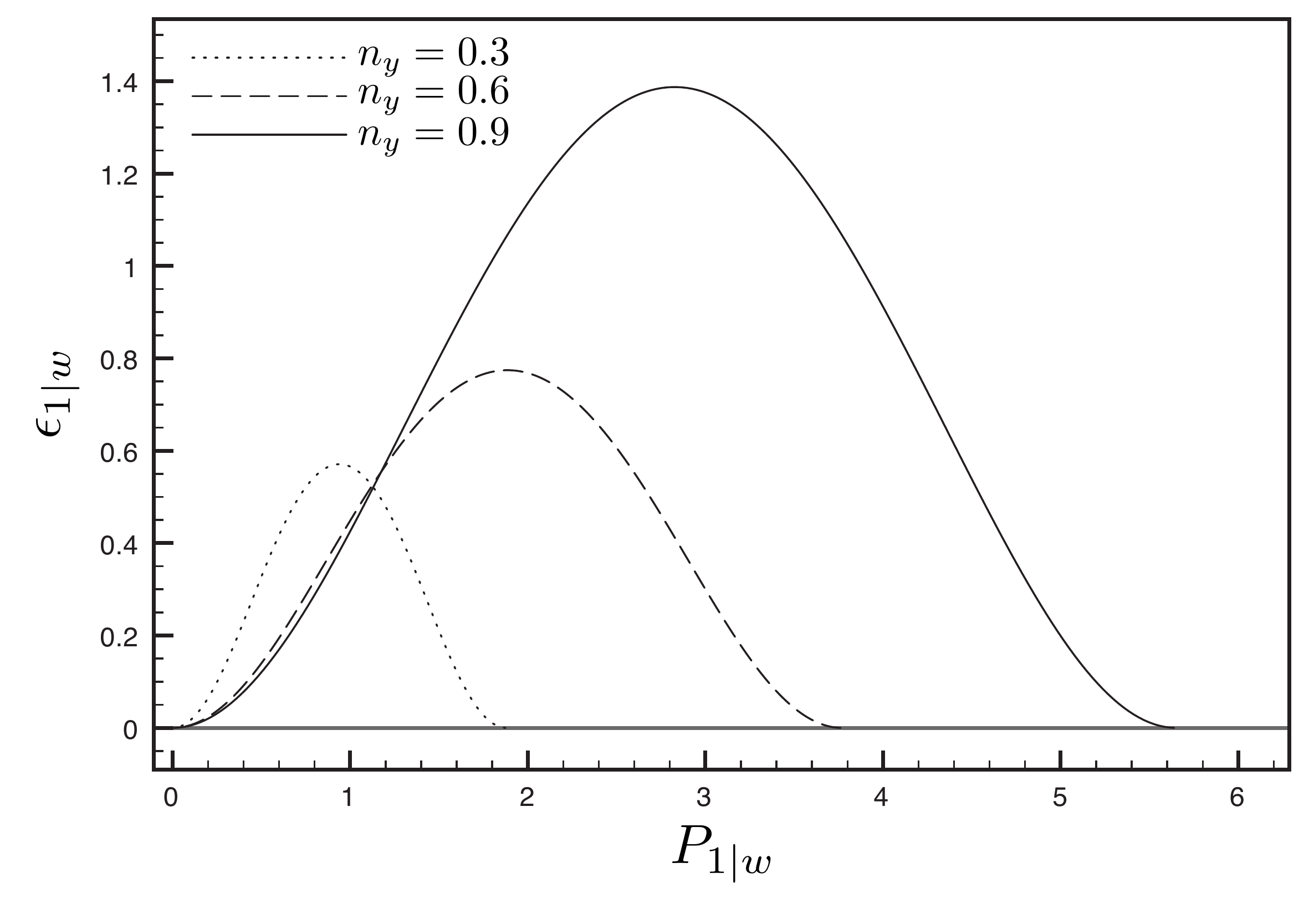}
\hfill
\includegraphics[width=0.48\linewidth]{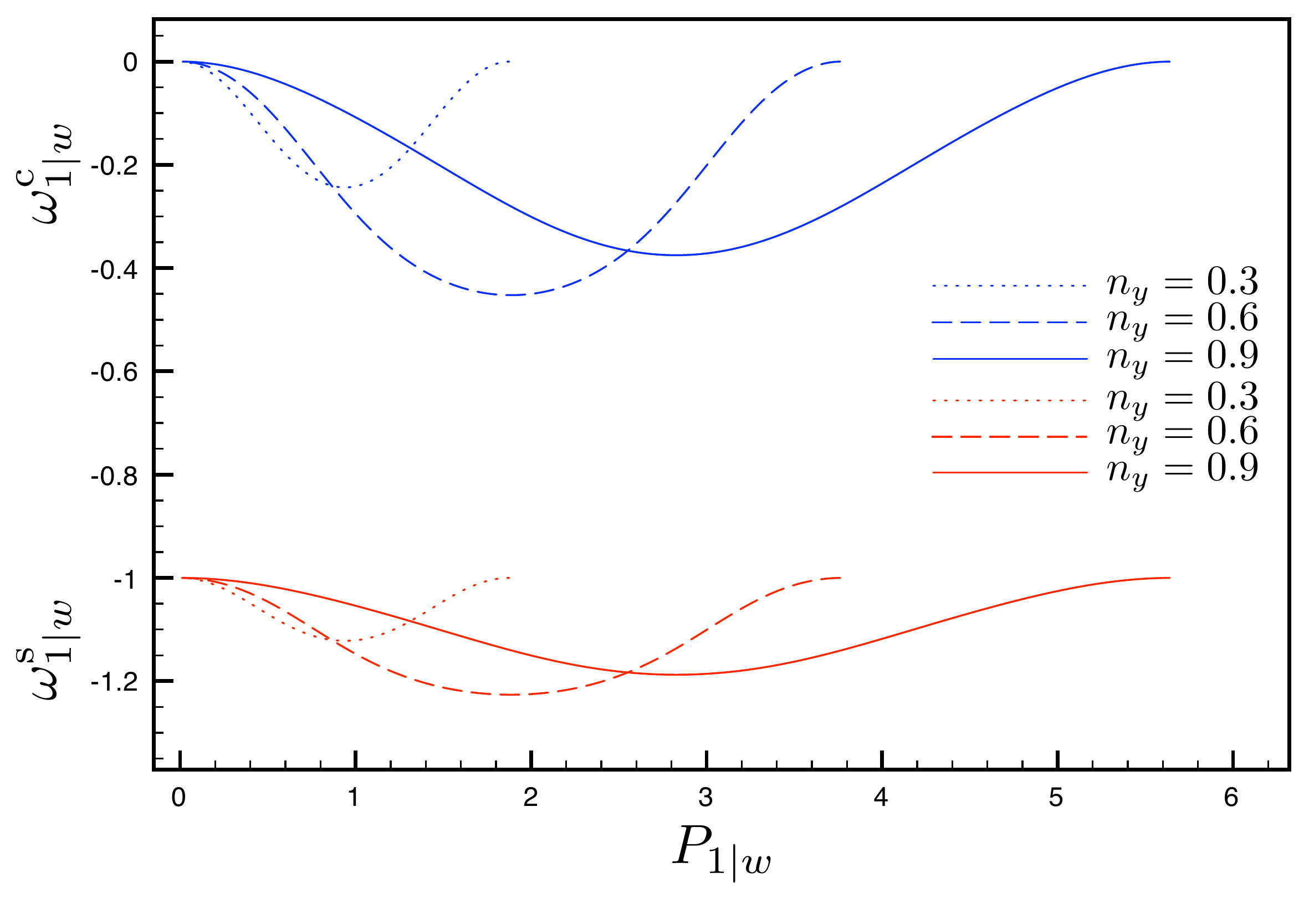}
\vspace{-0.5cm}
\caption{{\small   A-model: Plots of $\e_{1|w}(P_{1|w})$,  $\omega^{\rm c}_{1|w}(P_{1|w})$ and  $\omega^{\rm s}_{1|w}(P_{1|w})$  at $\uh=1$ and $B=0$.  }}
  \la{EPDscwA}
\end{figure}
The velocity can easily be read from the plots as it is the derivative of dressed energy with respect to dressed momentum.
Away from half-filling the dressed charge and spin of some strings gain dependence on rapidity. They are given by
\be
\begin{aligned}
\omega_y^{\rm c} = 1\,,&\quad \omega_{M|vw}^{\rm c} = 2 M -{\mathcal W}_M \,,\quad \omega_{M|w}^{\rm c} =-{\mathcal W}_M\,,\\
\omega_y^{\rm s} = {1\ov 2}\,,&\quad \omega_{M|vw}^{\rm s} = -{1\ov 2}{\mathcal W}_M \,,\quad \omega_{M|w}^{\rm s} =-M-{1\ov 2}{\mathcal W}_M\,,
\end{aligned}
\ee
where ${\mathcal W}_M(v)=-1\circledast_{Q_y}K_M$ is a non-negative function that goes to zero both at half-filling and  zero filling. We see that the $y$-particle behaves as an electron in this phase also, but that  it is now gapless. In $B=0$  magnetic field the $M|w$-strings  retain their quadratic dispersion away from half-filling but they are no longer pure spin, they gain charge opposite to that of an electron as  their energy increases. The dressed spin and charge of a $1|w$-string is plotted as a function of its dressed momentum for various fillings at $B=0$ and $\uh=1$  in Figure  \ref{EPDscwA}. In a $B>0$ magnetic field the $M|w$-strings become gapped. The $M|vw$-strings become dynamical in phase II but they are gapped throughout.

\subsection{B-model}\la{Bmodel}

Now we discuss the zero temperature excitations of the B-model. Let us begin again by recalling the ground state. The zero temperature TBA equations are
\be\la{BT0TBA}
\begin{aligned}
\e_y&=\E_y-\mu-B-\e_{1|vw}\star_{Q_{1|vw}}\, K_1\,,\\
\e_{M|vw}&=\E_{M|vw}-2M \mu + \e_y \circledast_{Q_y}\, K_M - \e_{1|vw}\star_{Q_{1|vw}}\, K_{1M}\,,\\
\e_{M|w}&=2MB + \e_y \circledast_{Q_y}\, K_M \,.
\end{aligned}
\ee
The dispersions $\E_y$ and $\E_{1|vw}$ are plotted as functions of rapidity $v$ in Figure \ref{EvwyB}.
\begin{figure}[t]
\includegraphics[width=0.48\linewidth]{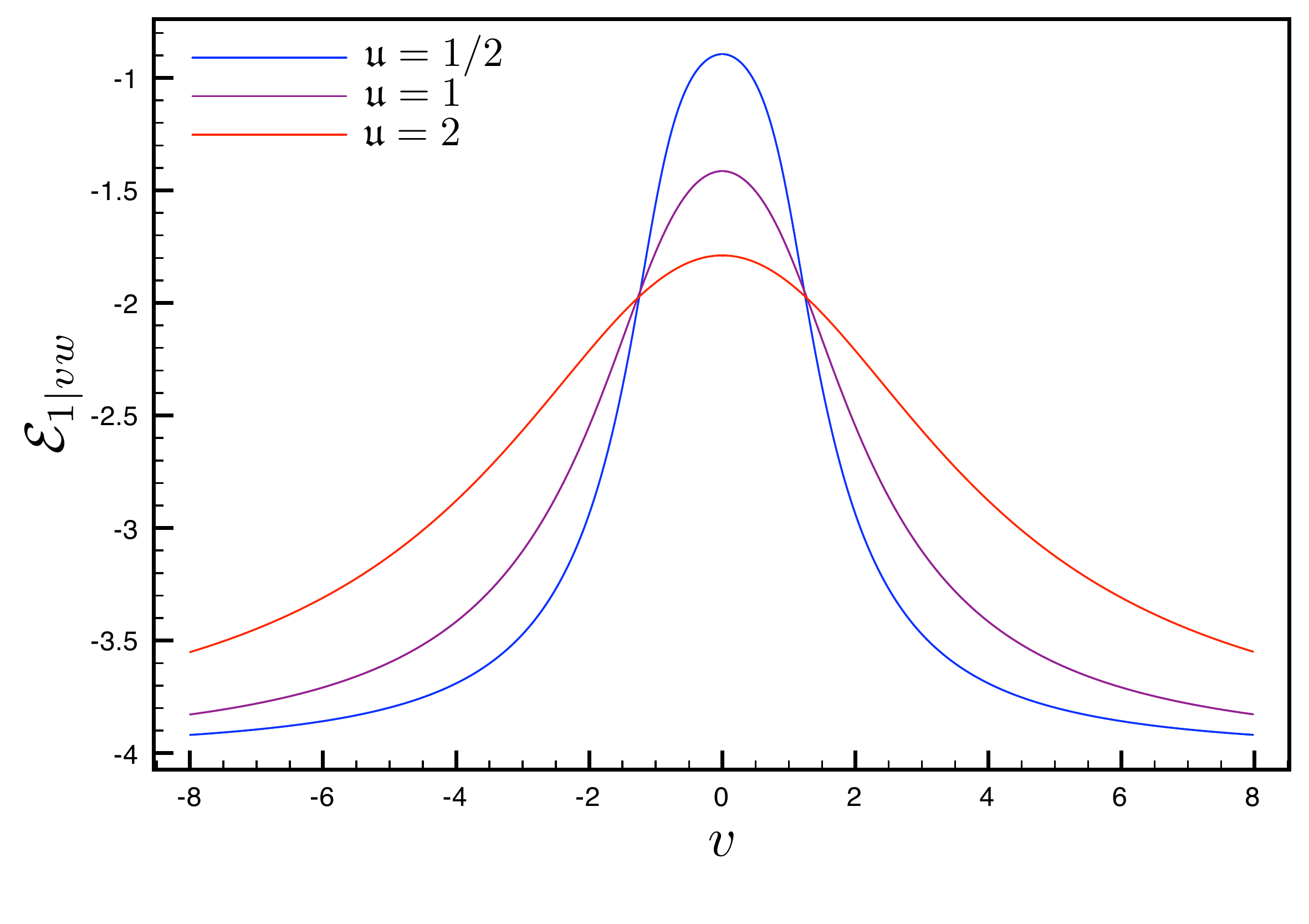}
\hfill
\includegraphics[width=0.48\linewidth]{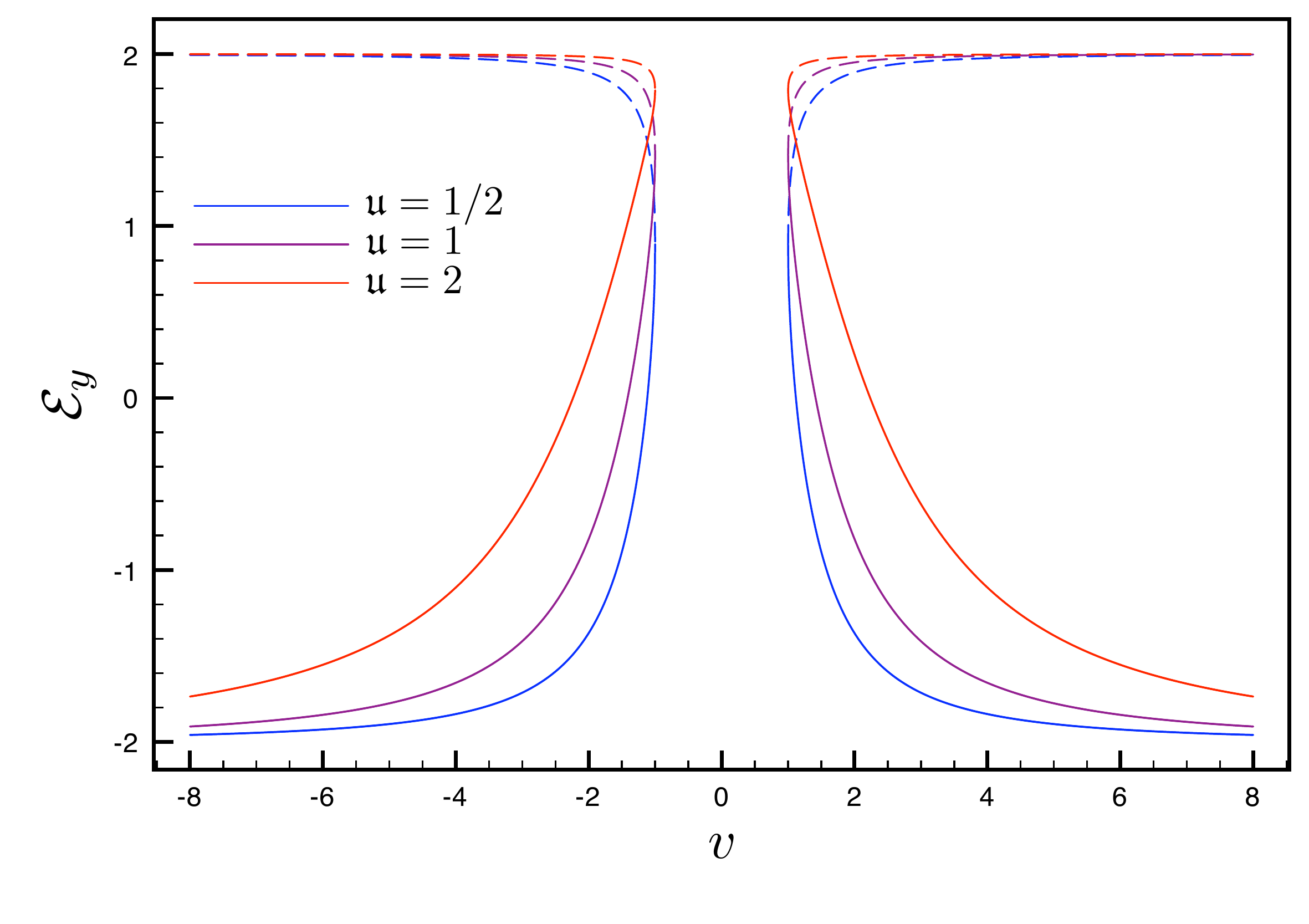}
\vspace{-0.5cm}
\caption{{\small Plots of $\E_{1|vw}(v)$ and $\E_{y}(v)$ for the B-model for $\uh=1/2,1,2$. In the plots of $\E_y(v)$ the $y_+$ branch is represented by a solid line and the $y_-$ branch is represented by a dashed line. }}
  \la{EvwyB}
\end{figure}
Here $1|vw$-strings and $y$-particles are the only strings that can have non-zero densities. The phase diagram is presented in Figure \ref{phasediagB}. 
For $B=0$ there exist only $1|vw$-strings and the ground state has zero magnetisation. In a $B>0$ magnetic field some of the bound pairs get broken introducing $y$-particles to the ground state.
From the TBA equations \eqref{BT0TBA} one can see that the $M|vw$-strings are type 2 strings while 
the $M|w$-strings are of type 1, and 
the $y$-particles should be treated as type 2 strings as discussed in appendix \ref{HSy}.
\begin{SCfigure}[2.6][t]
  \centering
  \includegraphics[width=0.23\textwidth]%
    {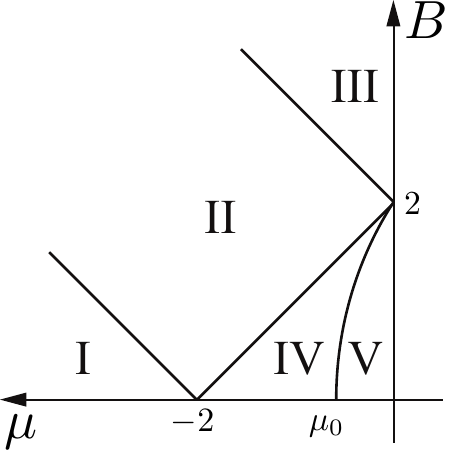}
    \hspace{1em}
  \caption{\small  B-model: Zero temperature phase diagram in the $\mu B$-plane. The phases identified are: I) empty band, II) partially filled and spin polarised band, III) half-filled and spin-polarised band, IV) partially filled and partially spin polarised band, V) half-filled and partially spin polarised band. The value $\mu_0$ ranges between 0 in the limit of weak coupling, and $-2+2\log 2\approx -0.6137$ in the limit of strong coupling. }
   \la{phasediagB}
\end{SCfigure}

\medskip

Let us first focus on excitations over the ground state when it is half-filled and has zero magnetisation. This is the subregion of phase V where $\mu_0\leq\mu\leq0$ and $B=0$.
 The TBA equations \ref{BT0TBA} can be solved explicitly with solution
\be\la{Bephf}
\e_{1|vw}=-\E_y\circledast s -\mu\,,\quad  \e_{M\geq2|vw}=-2(M-1)\mu \, , \quad  \e_y=\E_y-\E_{1|vw}\star s \,,\quad \e_{M|w}=0\,.
\ee
The $M|vw$-strings for $M\geq2$ and the $M|w$-strings are not dynamical. 
The $1|vw$-strings are half-filled and so we use the prescription \eqref{dNs} for handling excitations which change the range of mode numbers and get
\be\la{singr}
\delta N_{1|vw}^{\rm s}=-{1\ov 2}\delta N_{1|vw}^{\rm p}-{1\ov2}\delta N_y^{\rm p} - \sum_{M=2}^{\infty} \delta N_{M|vw}^{\rm p} \,.
\ee
Thus only excitations with $\delta N_{1|vw}^{\rm p}+\delta N_y^{\rm p}$ even are allowed, as $\delta N_{1|vw}^{\rm s}$ must be an integer.

 To calculate the dressed momenta let us make the branch choice  $b_{1|vw,M|vw}=1$ and $b_{1|vw,y}=0$, 
\be\la{BPhf}
\begin{aligned}
 P_{1|vw}&=-\pi+p_{1|vw}-{1\ov 2\pi} \Upsilon\star\Der{p_{1|vw}}{v}=- p_y \circledast s\,,\quad  P_{M\geq2|vw}=(M-1) \pi \mod 2\pi \, ,\\
P_y&=-\Psi+p_y- p_{1|vw}  \star s\,, \quad P_{M|w}=0\,,
\end{aligned}
\ee
where we have introduced the useful functions
\be
\begin{aligned}
\Upsilon(v)&=  \Theta_1\star s(v) = i\log\Big[{\Gamma(\frac{1}{2}+i \frac{v}{4 \uh}) \Gamma(1-i \frac{v}{4 \uh}) \ov  
 	\Gamma(\frac{1}{2}-i \frac{v}{4 \uh}) \Gamma(1 + i \frac{v}{4 \uh}) }\Big] \,, \\
 \Psi(v)&=\Theta_2\star s(v)-\Theta_1(v) = {\pi\ov 2} - 2 \arctan \Big[ \exp \big( {\pi v\ov 2 \uh} \big)\Big]\,.
\end{aligned}
\ee  
The range of $P_{1|vw}$ is $(-{\pi\ov2},{\pi\ov 2})$ while the range of $P_y$ is $(-{3\pi\ov2},-{\pi\ov2})\cup({\pi\ov 2},{3\pi\ov2})$. Let us remark that when taken modulo $2\pi$ the range of $P_y$ will have an overlap.
The singular $1|vw$-strings appearing through eq. \eqref{singr} have rapidity $v^{\rm max}=0$ and so carry momentum $0$. The dressed charge and spin are
\be\la{wschfB}
\begin{aligned}
\omega_y^{\rm c} = 0\,,&\quad \omega_{1|vw}^{\rm c} = 1\,,\quad \omega_{M\geq2|vw}^{\rm c} = 2M-2 \,,\quad \omega_{M|w}^{\rm c} =0\,,\\
\omega_y^{\rm s} = {1\ov 2}\,,&\quad \omega_{M|vw}^{\rm s} = 0 \,,\quad \omega_{M|w}^{\rm s} =-M\,,
\end{aligned}
\ee
 and we observe that the excitations are spin-charge separated in this phase. The removed $1|vw$-strings get dressed as 
holons while added $y$-particles get dressed as
 spinons. The energies and momenta of the holons, antiholons and spinons are
\be
\begin{aligned}
E_{\rm h} &= -\e_{1|vw}=\E_y\circledast s +\mu\,, & P_{\rm h} &= - P_{1|vw}=p_y\circledast s \,,\\
E_{\rm \bar h} &= -\e_{1|vw}+\e_{2|vw}=\E_y\circledast s -\mu\,, & P_{\rm \bar h} &= - P_{1|vw} +P_{2|vw} =-\pi\,\mbox{sign}+p_y\circledast s\,,\\
E_{\rm s} &= E_{\rm \bar s}=\e_y=\E_y-\E_{1|vw}\star s\,, & P_{\rm s} &= P_{\rm \bar s} = P_y =-\Psi+p_y-p_{1|vw}\star s \,.
\end{aligned}
\ee
Here an antiholon is identified as a composite excitation of a holon and a $2|vw$-string because a $2|vw$-string is not dynamical. Let us remark that through
 eqs. \eqref{Bephf}, \eqref{BPhf} and \eqref{wschfB} it can be seen that its addition can also be regarded as the action of the  charge $\su(2)$ raising operator on a state. Similarly the spinon ${ \bar s}$ is a composite of a spinon ${ s}$ and a  $1|w$-string.
Plots of $E_{\rm h}(P_{\rm h})$ and $E_{\rm s}(P_{\rm s})$ are given in Figure \ref{EPhsB} for various values of $\uh$. 
\begin{figure}[t]
\includegraphics[width=0.48\linewidth]{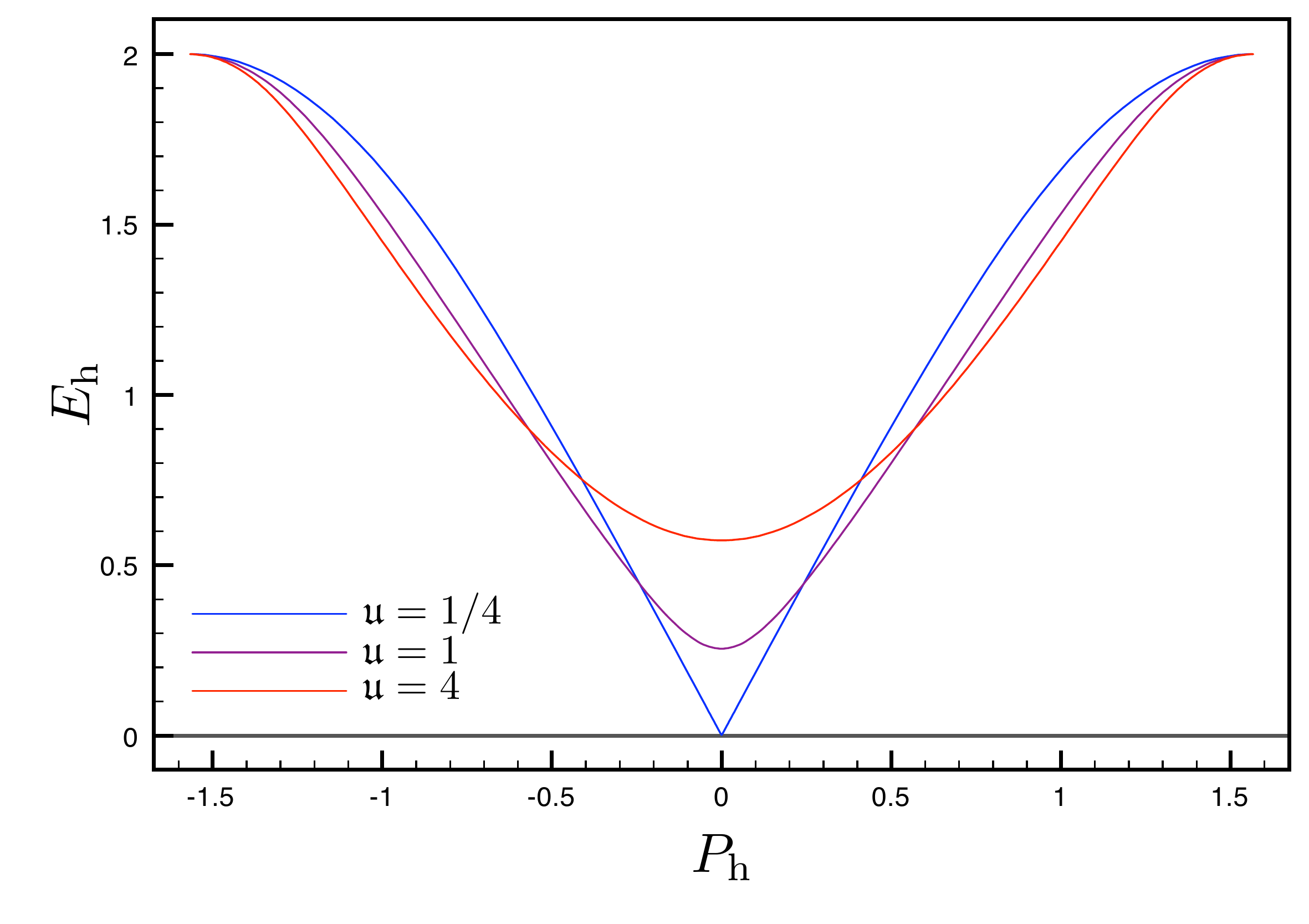}
\hfill
\includegraphics[width=0.48\linewidth]{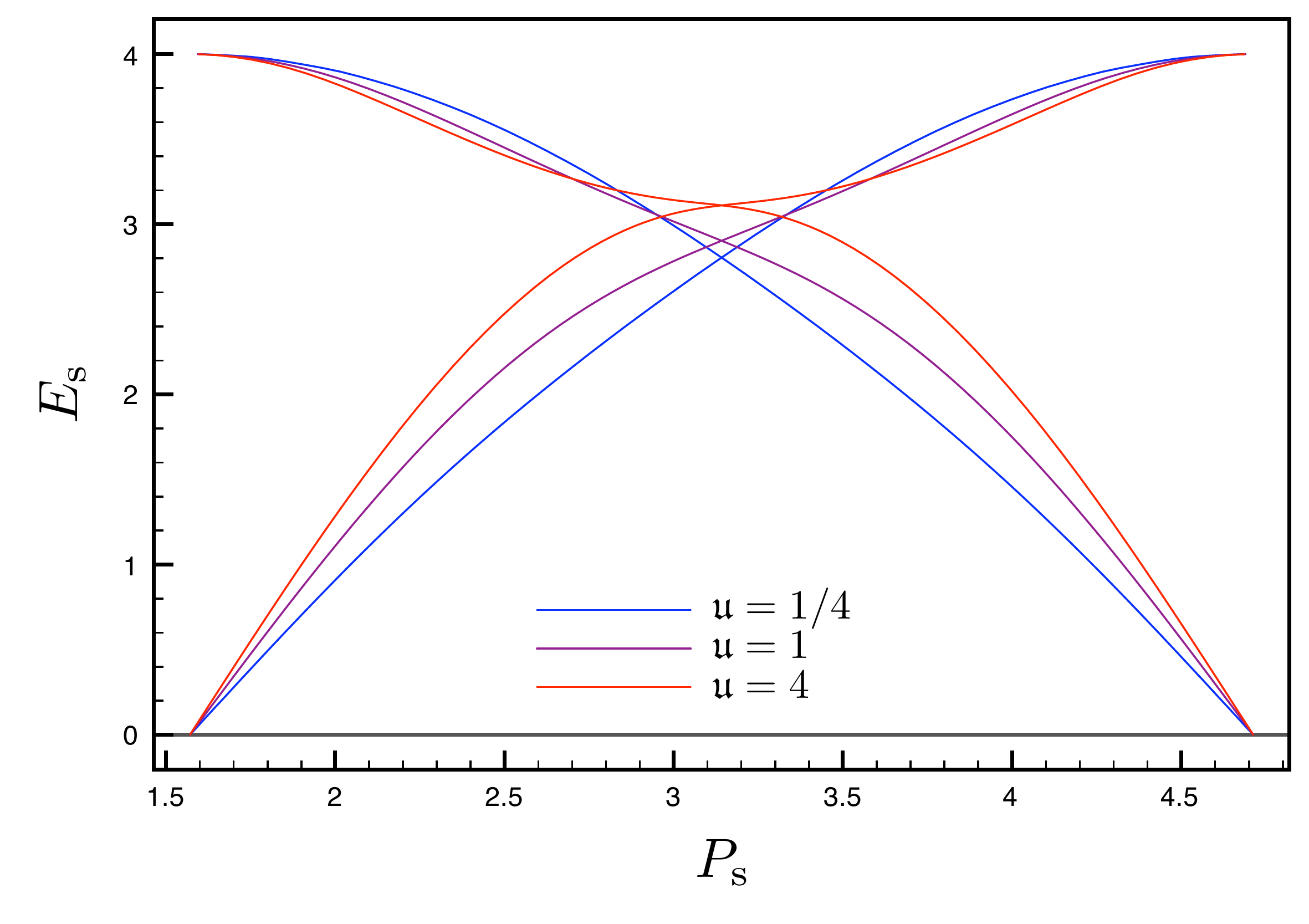}
\vspace{-0.5cm}
\caption{{\small Plots of $E_{\rm h}(P_{\rm h})$ and $E_{\rm s}(P_{\rm s})$ for the half-filled B-model  at $\mu=0$, $B=0$. }}
  \la{EPhsB}
\end{figure}
 The velocities can again easily be read from the slopes.
The holons are gapped for $\mu>\mu_0$. The gap goes to zero in the weak coupling $\uh\to0$ limit while the gap has a maximal value of $2-2\log 2\approx 0.6137$ at $\mu=0$ in the strong coupling $\uh\to\infty$ limit.
 The spinons are gapless and display an ``hourglass''  dispersion. The similarity to experimental data on spinon scattering in some cuprate materials, say Fig. 2 of \cite{spinstudies}, cannot go unremarked. In the strong coupling $\uh\to\infty$ limit the two lower wings join differentiably at ${\pi}$ and the upper wings can be understood as the contributions of dressed electrons on doubly occupied sites \cite{FQ11}.

Next let us calculate the dressed scattering phases to examine the scattering of the holons and spinons. 
First we present explicitly the bare scattering phases
\be\notag
\begin{aligned}
\phi_{1|vw,1|vw}(v,t)& =2\pi-\pi\,\mbox{sign}(v)+\Theta_2(v-t) \,,\quad \\
\phi_{1|vw,M\geq2|vw}(v,t) &= 2\pi -2\pi\,\mbox{sign}(v)+\Theta_{1M}(v-t)\,,\\
\phi_{1|vw,y}(v,t)&=\phi_{y,1|vw}(v,t)=-\pi\,\mbox{sign}(v)+\Theta_1(v-t)
 \,,\\
\phi_{1|vw,M|w}(v,t)&=0\,,\quad \phi_{y,y}(v,t)=0 \,,\quad 
\phi_{y,M|w}(v,t)=-\pi\,\mbox{sign}(v)+\Theta_M(v-t) \,.
\end{aligned}
\ee
To calculate the dressed scattering it will be necessary to redistribute the contributions of the singular strings as in eq. \eqref{redist}. 
Solving eqs. \eqref{eqPhi} for $\Phi_{1|vw,1|vw}$, $\Phi_{M|vw,1|vw}$ and  $\Phi_{y,1|vw}$  we get
\be\la{phi0}
\begin{aligned}
\Phi_{1|vw,1|vw}(v,t)&=\pi-\pi\,\mbox{sign}(v)+\Upsilon(v) +\Upsilon(v-t) \,,\\
\Phi_{M\geq2|vw,1|vw}(v,t)&=-2\pi\,\mbox{sign}(v)+\Theta_{M-1}(v) +\Theta_{M-1}(v-t) \,,\\
\Phi_{y,1|vw}(v,t)&=-\pi-\pi\,\mbox{sign}(v)-\Psi(v) -\Psi(v-t) \,,
\end{aligned}
\ee
where the identity $\pi\,\mbox{sign}\star K_1=\Theta_1$ has been used.
As we are interested in the scattering phase shift, which is defined modulo $2\pi$, $\mbox{sign}(v)$ can be dropped from \eqref{phi0}. Taking into account the scattering phases of singular strings the dressed scattering phases are 
\be\notag
\begin{aligned}
&
\begin{aligned}
\Phi^{\rm h.f.}_{1|vw,1|vw}(v,t)&=\Upsilon(v-t) \,,\quad& \Phi^{\rm h.f.}_{y,y}(v,t)&=\pi - \Upsilon(v-t)\,,\\
\Phi^{\rm h.f.}_{\a\beta}(v,t)&=\Phi^{\rm h.f.}_{\beta\a}(v,t)\,,\quad &\Phi^{\rm h.f.}_{y,1|vw}(v,t)&=\pi+\Psi(v-t)\,,\\
\Phi^{\rm h.f.}_{1|vw,M\geq2|vw}(v,t)&=\pi +\Theta_{M-1}(v-t) \,,\quad &\Phi^{\rm h.f.}_{y,M|w}(v,t)&=\pi+\Theta_M(v-t)\,,\\
\Phi^{\rm h.f.}_{M|vw,N|w}(v,t)&=0\,,\quad
&\Phi^{\rm h.f.}_{y,M\geq2|vw}(v,t)&=0 \,,
\end{aligned}\\
&\Phi^{\rm h.f.}_{M\geq2|vw,N\geq2|vw}(v,t)=\delta_{MN}\pi  +\Theta_{MN}(v-t) -\Theta_{M+N}(v-t) -\Theta_{M+N-2}(v-t)\,,\\
& \Phi^{\rm h.f.}_{M|w,N|w}(v,t)= \delta_{MN}\pi  -\Theta_{MN}(v-t)\,.
\end{aligned}
\ee
Let us compute explicitly the phase shifts for the charge triplet and charge singlet excitations. 
\newline\smallskip\noindent{\it Charge triplet}: holon-holon scattering. Here  two $1|vw$-strings
with rapidities $v_1$ and $v_2$  are removed.
 Let us say that $v_1$ has a greater velocity $\Der{\e_{\rm h}}{P_{\rm h}}$ than $v_2$, and let us denote this as $v_1\succ v_2$. 
 Then 
\be\notag
 F_{1|vw}(v)= -\Phi^{\rm h.f.}_{1|vw,1|vw}(v,v_1)-\Phi^{\rm h.f.}_{1|vw,1|vw}(v,v_2)\,
\ee
and the phase shift is
\be 
\delta_{\rm CT} =\pi+ \Upsilon(v_1-v_2)\,.
\ee
{\it Charge singlet}:  holon-antiholon scattering. Here two $1|vw$-strings
with rapidities $v_1\succ v_2$ are removed and 
 a $2|vw$-string with rapidity $\tilde v$ is added. 
The rapidity $\tilde v$ of the added $2|vw$-string can be 
fixed through eq. \eqref{shifts1} using $P_{2|vw}=0$,
\begin{align}\notag
F_{2|vw}& = -\Phi^{\rm h.f.}_{2|vw,1|vw}(v,v_1)-\Phi^{\rm h.f.}_{2|vw,1|vw}(v-v_2) + \Phi^{\rm h.f.}_{2|vw,2|vw}(v-\tilde v)\\\notag
	& = \pi - \Theta_1(v-v_1) - \Theta_1(v-v_2) + \Theta_2(v-\tilde v)\,,
\end{align}
and so $F_{2|vw}(\tilde v)=\pi$ gives $\tilde v={v_1+v_2\ov 2}$. 
Here
\be\notag
 F_{1|vw}(v)= -\Phi^{\rm h.f.}_{1|vw,1|vw}(v,v_1)-\Phi^{\rm h.f.}_{1|vw,1|vw}(v,v_2) + \Phi^{\rm h.f.}_{1|vw,2|vw}\big(v,{v_1+v_2\ov 2}\big)\,
\ee
and  thus
\be 
\delta_{\rm CS} =  \Upsilon(v_1-v_2) - \Theta_1\big({v_1-v_2\ov 2}\big) \,.
\ee
Let us remark that these results as functions of the rapidity are the same as those of the half-filled Hubbard model, see e.g. eqs. (7.124) and (7.126) of \cite{bookH}. The scattering shifts for the spin triplet, singlet and spin-charge excitations can be computed similarly and also agree with those of the Hubbard model, eqs. (7.139), (7.141) and eq. (7.145)  of \cite{bookH}.

\medskip

Now let us consider the less than half-filled phase while still keeping $B=0$. In Figure \ref{phasediagB} this is the portion of phase IV along the $\mu$-axis. The magnetisation is zero and the filling is  $2n_{1|vw}$.
Here again $\e_{M|w}=0$ and $P_{M|w}=0$ for $M|w$-strings but to find the dressed energies and momenta of $M|vw$-strings and $y$-particles one must solve the TBA equations \eqref{BT0TBA} numerically. 
Taking all $b_{\a\beta}=0$, the dressed momentum  for $y$-particles takes values in $\pi n_{1|vw}<|P_y|< \pi+\pi n_{1|vw}$, for $1|vw$-strings  in $\pi n_{1|vw}<|P_{1|vw}|< \pi$,  for $M|vw$-strings with $M\geq3$ odd in $2\pi n_{1|vw}<|P_{M|vw}|< \pi$, and  for $M|vw$-strings with $M$ even in $\pi+2\pi n_{1|vw}<|P_{M|vw}|<  2\pi$. These ranges should be considered modulo $2\pi$ but it is more convenient in plots to use the ranges specified here.  The other branches are obtained by shifts of $2\pi n_{1|vw}$.
\begin{figure}[t]
\includegraphics[width=0.48\linewidth]{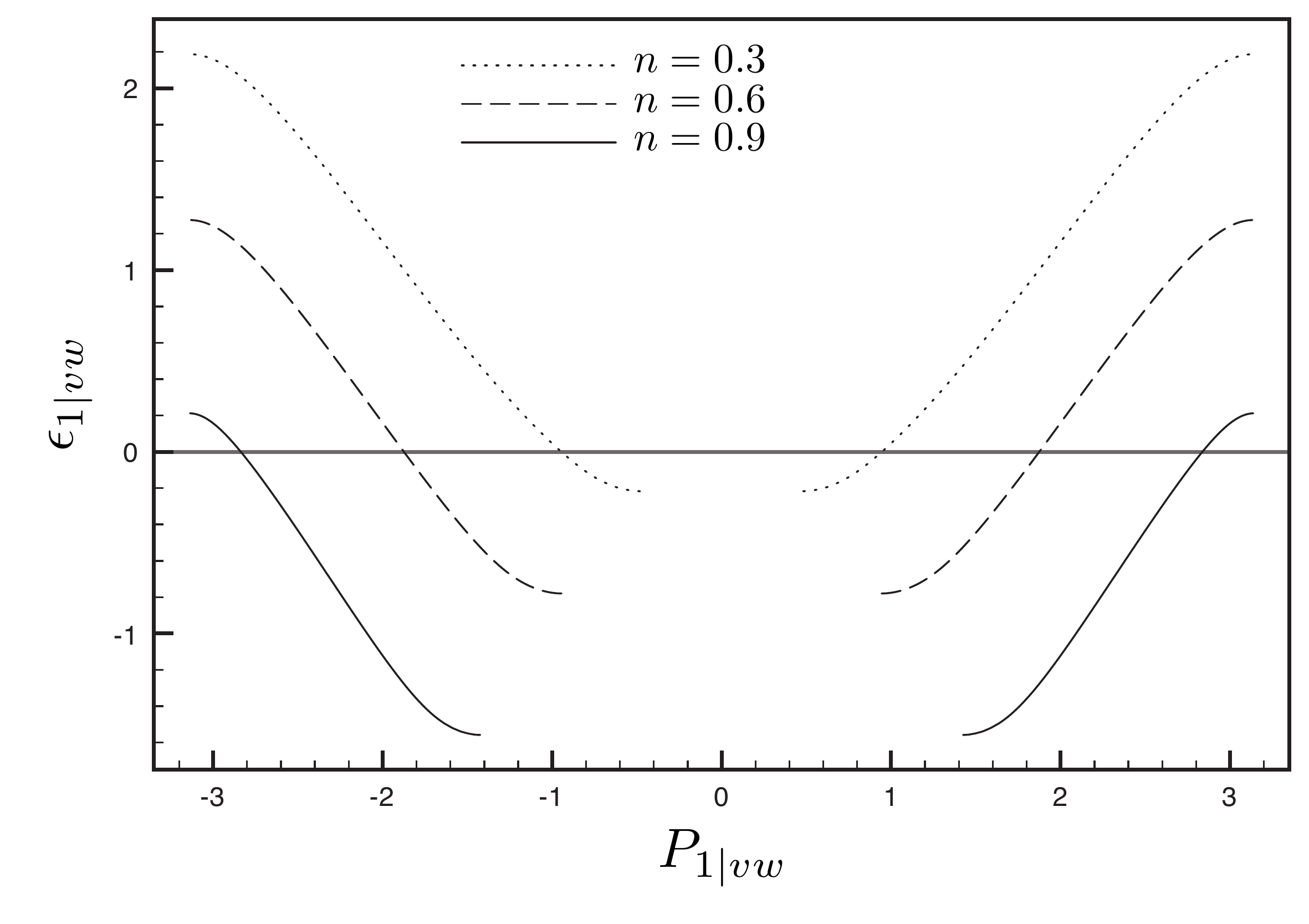}
\hfill
\includegraphics[width=0.48\linewidth]{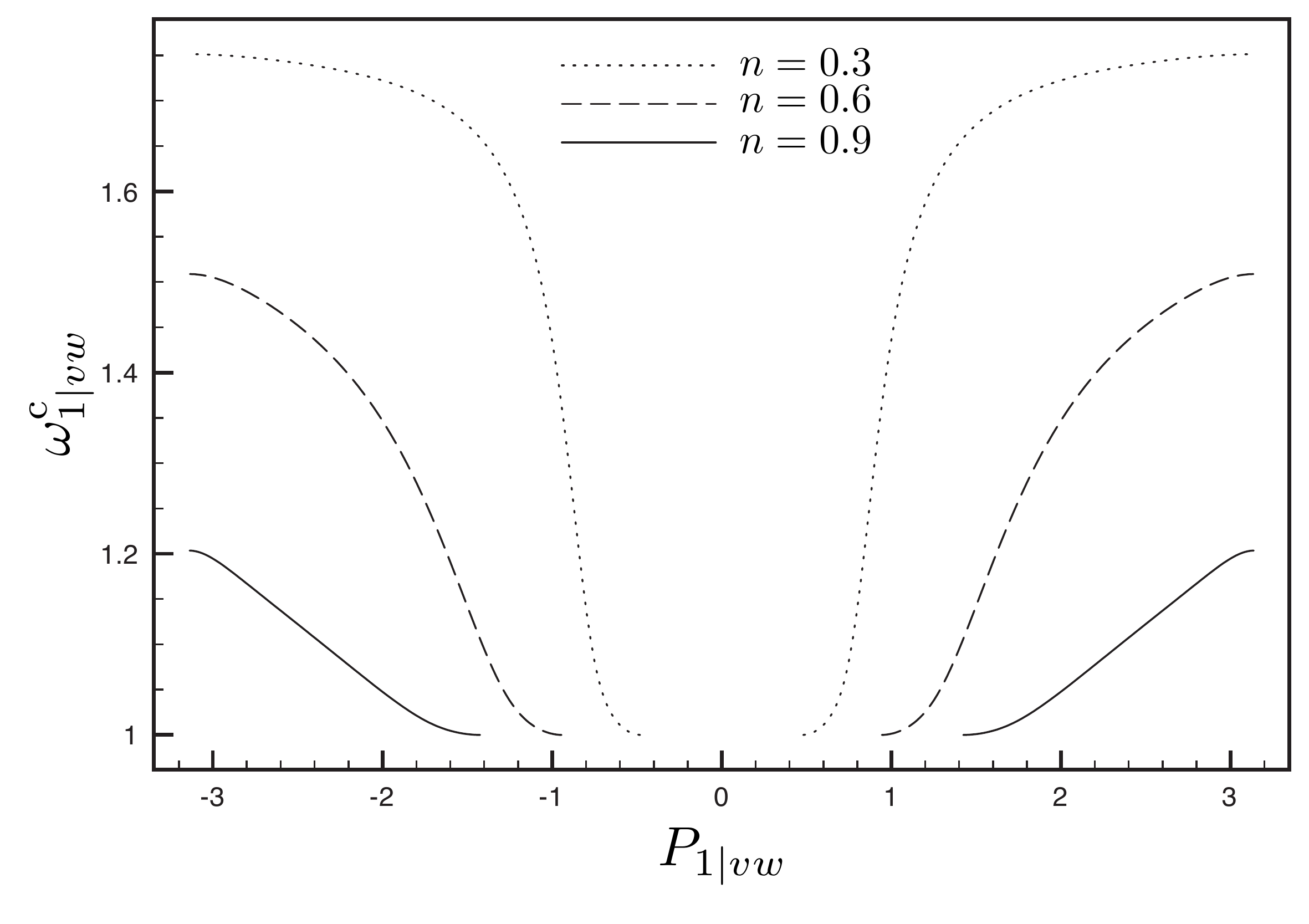}
\vspace{-0.5cm}
\caption{{\small  B-model: Plots of dressed energy $\e_{1|vw}(P_{1|vw})$ and dressed charge  $\omega^{\rm c}_{1|vw}(P_{1|vw})$ at $\uh=1$ and $B=0$ for various fillings.  }}
  \la{ECvwBlhf}
\end{figure}
The dressed spins of the excited strings take their bare values while the equations for dressed charge are
\be
 \omega_y^{\rm c} = 1 -  \omega_{1|vw}^{\rm c}\star_{Q_{1|vw}}K_1\,,\quad
 \omega_{M|vw}^{\rm c} = 2M -  \omega_{1|vw}^{\rm c}\star_{Q_{1|vw}}K_{1M}\,,\quad 
 \omega_{M|w}^{\rm c} =0\,.
 \ee
These are rapidity dependent for the $y$-particles and $M|vw$-strings. Let us remark however that at $v=\pm\infty$ the dressed charges take the values they have at half-filling \eqref{wschfB}
\be
\omega_y^{\rm c}(\pm\infty)=0\,,\quad \omega_{1|vw}^{\rm c}(\pm\infty)=1\,,\quad \omega_{M\geq2|vw}^{\rm c}(\pm\infty)=2M-2\,.
\ee
In Figure \ref{ECvwBlhf}
  the dressed energy and dressed charge of a $1|vw$-string as function of its dressed momentum is plotted for various fillings at $\uh=1$. The corresponding plots for $y$-particles are given in Figure \ref{ECyBlhf}  and the ``hourglass'' behaviour about $\pi$ is seen again. 
  \begin{figure}[t]
\includegraphics[width=0.48\linewidth]{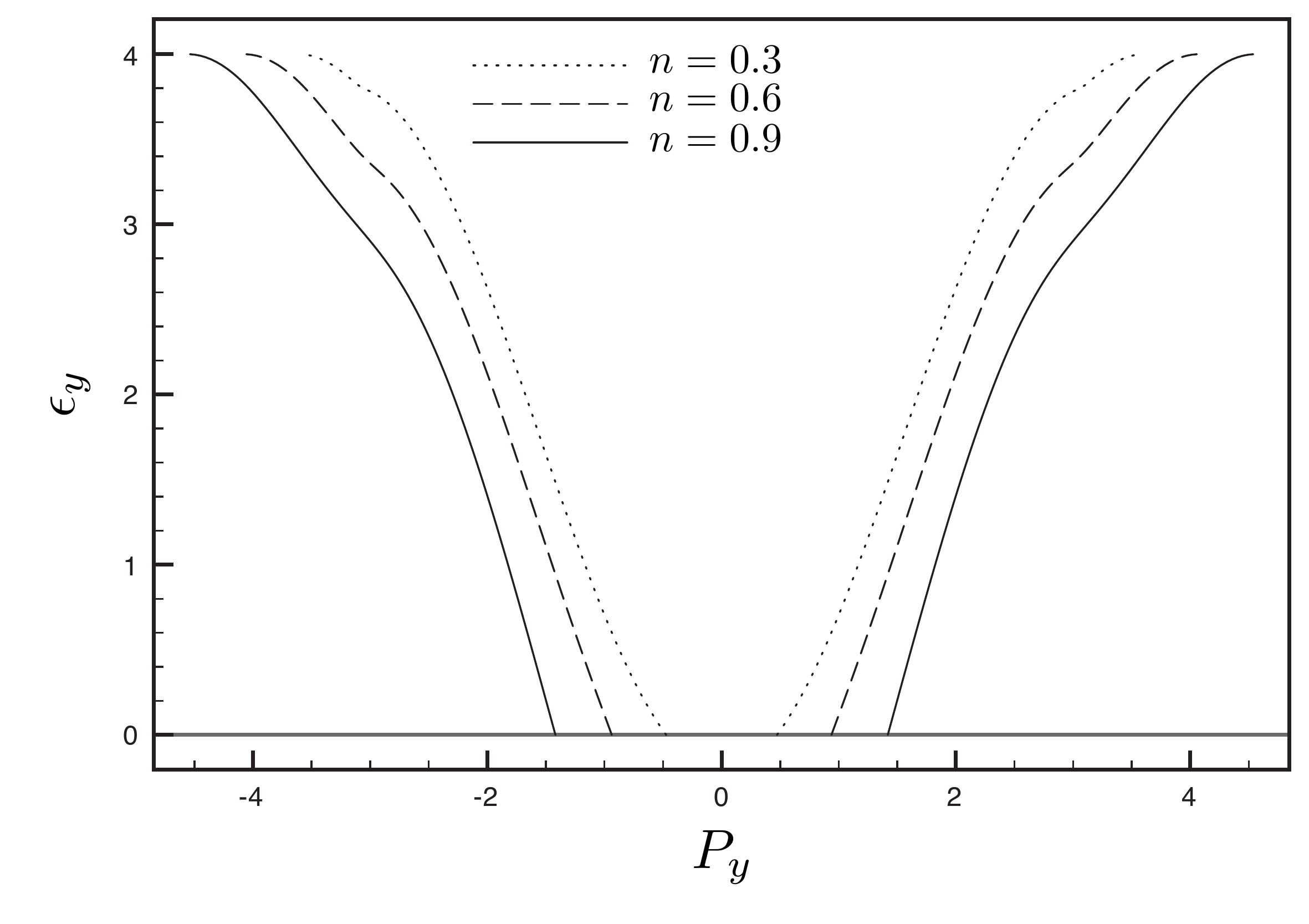}
\hfill
\includegraphics[width=0.48\linewidth]{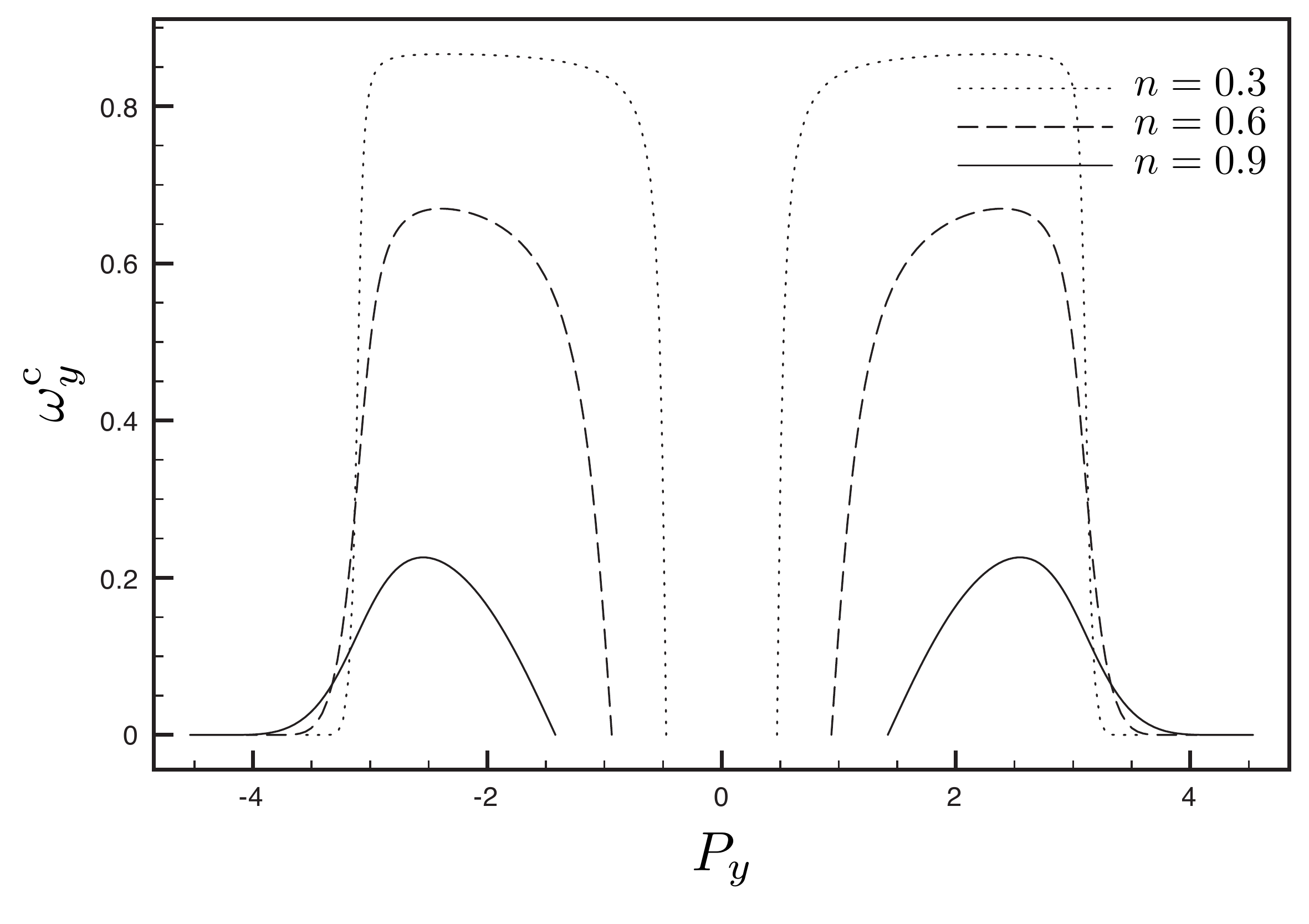}
\vspace{-0.5cm}
\caption{{\small  B-model: Plots of dressed energy $\e_{y}(P_{y})$ and dressed charge  $\omega^{\rm c}_{y}(P_{y})$ at $\uh=1$ and $B=0$ for various fillings.  }}
  \la{ECyBlhf}
\end{figure}
  Both excitations are gapless. The $M|vw$-strings with $M\geq2$ are dynamical but have a gap of $-2(M-1)\mu$. 
  The $1|vw$-strings remain spinless but $y$-particles with non-zero energy here have dressed charge in addition to their spin. 
  At low energies the charge carried by a $y$-particle scales with its energy, and moreover the magnitude of the charge carried increases    sharply as the filling is decreased. 
  Thus the excited quasi-particles are not spin-charge separated away from half-filling. Nevertheless, in the limit of zero energy the  quasi-particles carry either spin or charge and so this is compatible with spin-charge separated wave-like excitations that one may expect in the continuum limit, as in the Luttinger liquid. Let us remark that, as can be seen from Figure \ref{ECvwBlhf}, the charge of a zero-energy $1|vw$-string becomes greater than 1 at less than half-filling and thus we expect that the charge carried by a charge-wave gets increased at reduced filling.

It is noteworthy that at $B=0$ one can clearly see that the $1|vw$-string is a spin-singlet bound state. Let us show this. 
The spin singlet excitation is achieved by adding two $y$-particles with rapidities $v_1$ and $v_2$, and adding
 a $1|w$-string with rapidity $w$ which we initially take to be arbitrary. 
The relevant dressed phase shifts are
\begin{align}\notag
\Phi_{y,y}(v,t)&=-\big(K_1 \star_{Q_{1|vw}}\Phi_{1|vw,y}\big)(v,t)\,,\\\notag
 \Phi_{1|vw,y}(v,t)&=-\pi\,\mbox{sign}(v)+\Theta_1(v,t)- \big(K_2 \star_{Q_{1|vw}}\Phi_{1|vw,y}\big)(v,t)\,,\\\notag
\Phi_{y,1|w}(v,t)&=\pi+\Theta_1(v-t) \,,\, \Phi_{1|w,y}(v,t)=\pi+\Theta_1(v-t) \,,\, \Phi_{1|w,1|w}(v,t)=\pi-\Theta_2(v-t) \,.
\end{align}
The rapidity $w$  is fixed to $w={v_1+v_2\ov 2}$ by eq. \eqref{shifts1} as $P_{1|w}=0$. The scattering phase shift is
\be
F_{y}(v_1) = \Phi_{y,y}(v_1,v_1)+\Phi_{y,y}(v_1,v_2)+\pi+\Theta_1\big({v_1-v_2\ov 2}\big)\,.
\ee
Note that $\Theta_1(\pm i\uh)=\pm i\infty$ and so the final term gives rise to a pole of the S-matrix at $v_1=v_2-2i\uh$. Although the first two terms cannot be obtained explicitly it can be seen analytically that they cancel the pole through the term $\Theta_1$ in $\Phi_{1|vw,y}$ for $v\in Q_{1|vw}$, while for  $v\notin Q_{1|vw}$ the pole remains.  The pole corresponds to a bound state of a $y_-$- and a $y_+$-particle as ${\rm Im}\, P_-(v)>0$ for ${\rm Im}\, v\neq0$ and  ${\rm Im}\, P_+(v)<0$ for ${\rm Im}\, v\neq0$.
The bound state, with rapidity $v$, thus corresponds to a spin singlet excitation with 
\be
v_1=v-i \uh\,,\quad v_2=v+i \uh\,,\quad \tilde w=v\,.
\ee
and the changes of energy and momentum are
\be\la{BSSb}
\Delta E  = \e_{1|vw}(v)  \,,\quad
\Delta P  = P_{1|vw}(v)	\qquad\mbox{for } v\notin Q_{1|vw} \,.
\ee
Here the identity $K_1(v+i\uh- i0)+K_1(v-i\uh+i0)=\delta(v)+K_2(v)$ was used to obtain the energy and eq. \eqref{eqP} was used to obtain the momentum.
Hence the bound state is indeed an added $1|vw$-string. This indicates that spin-spin interactions are responsible for the pairing of the electrons into $1|vw$-strings.

\medskip

 Let us consider briefly the effect of a $B>0$ magnetic field at half-filling, the interior of phase V in Figure \ref{phasediagB}. Here there are both $1|vw$-strings and $y$-particles in the ground state and it has a magnetisation between 0 and $1/2$. The $1|vw$-strings are at half-filling\footnote{There are no holes for $1|vw$-strings here. There are less of them than there are at $B=0$ as their range of mode numbers is decreased by the presence of the $y$-particles.} and so excitations must satisfy \eqref{singr}, that is, only excitations with $\delta N_{1|vw}^{\rm p}+\delta N_y^{\rm p}$ even are allowed. 
The $1|vw$-strings are gapped and have dressed charge 1 while the $y$-particles have dressed charge zero and are gapless.  The $M\geq2|vw$-strings are non-dynamical while the $M|w$-strings are dynamical but gapped.
The spin dressing equations for $y$-particles and  $1|vw$-strings are
  \begin{figure}[t]
\includegraphics[width=0.48\linewidth]{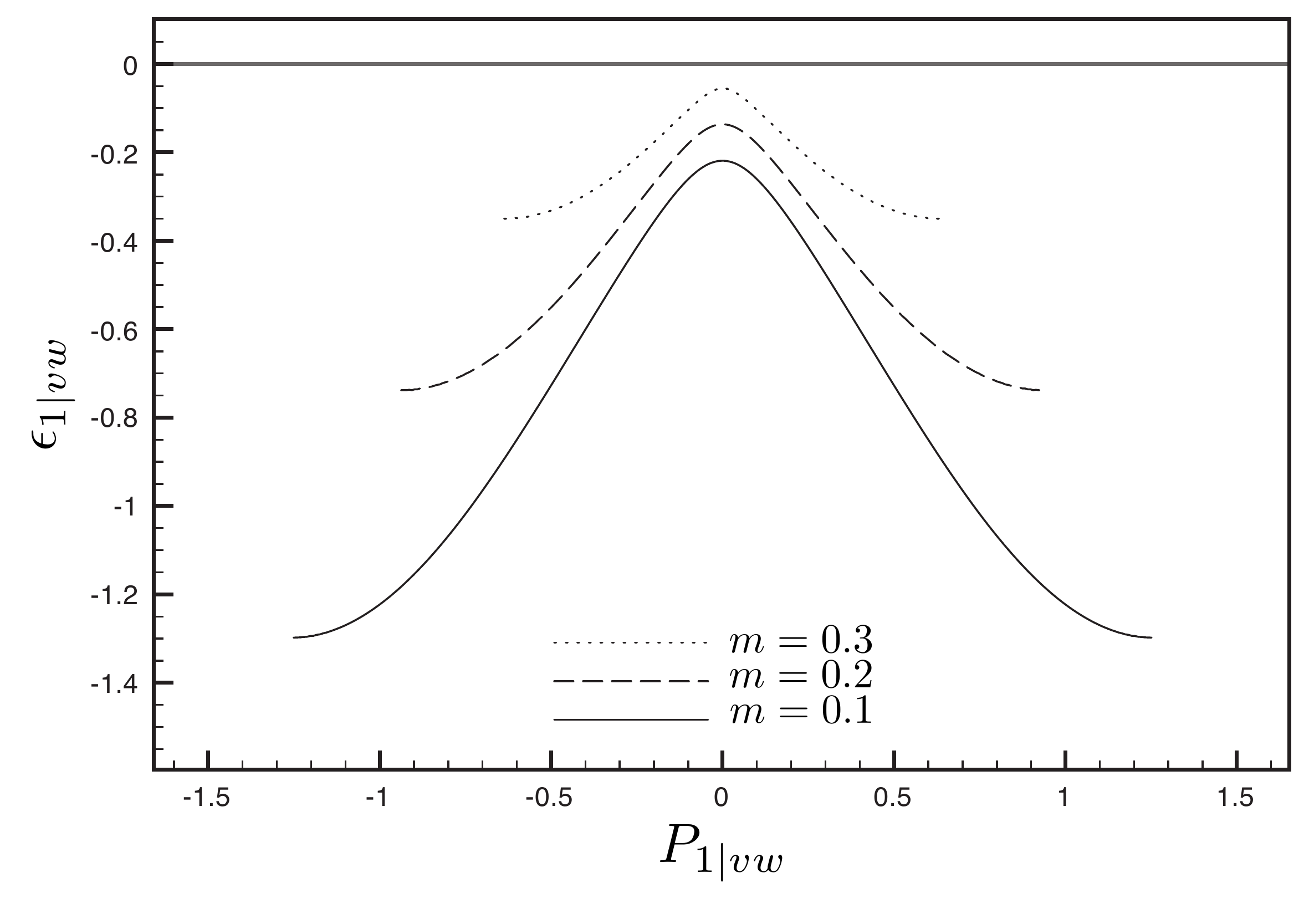}
\hfill
\includegraphics[width=0.48\linewidth]{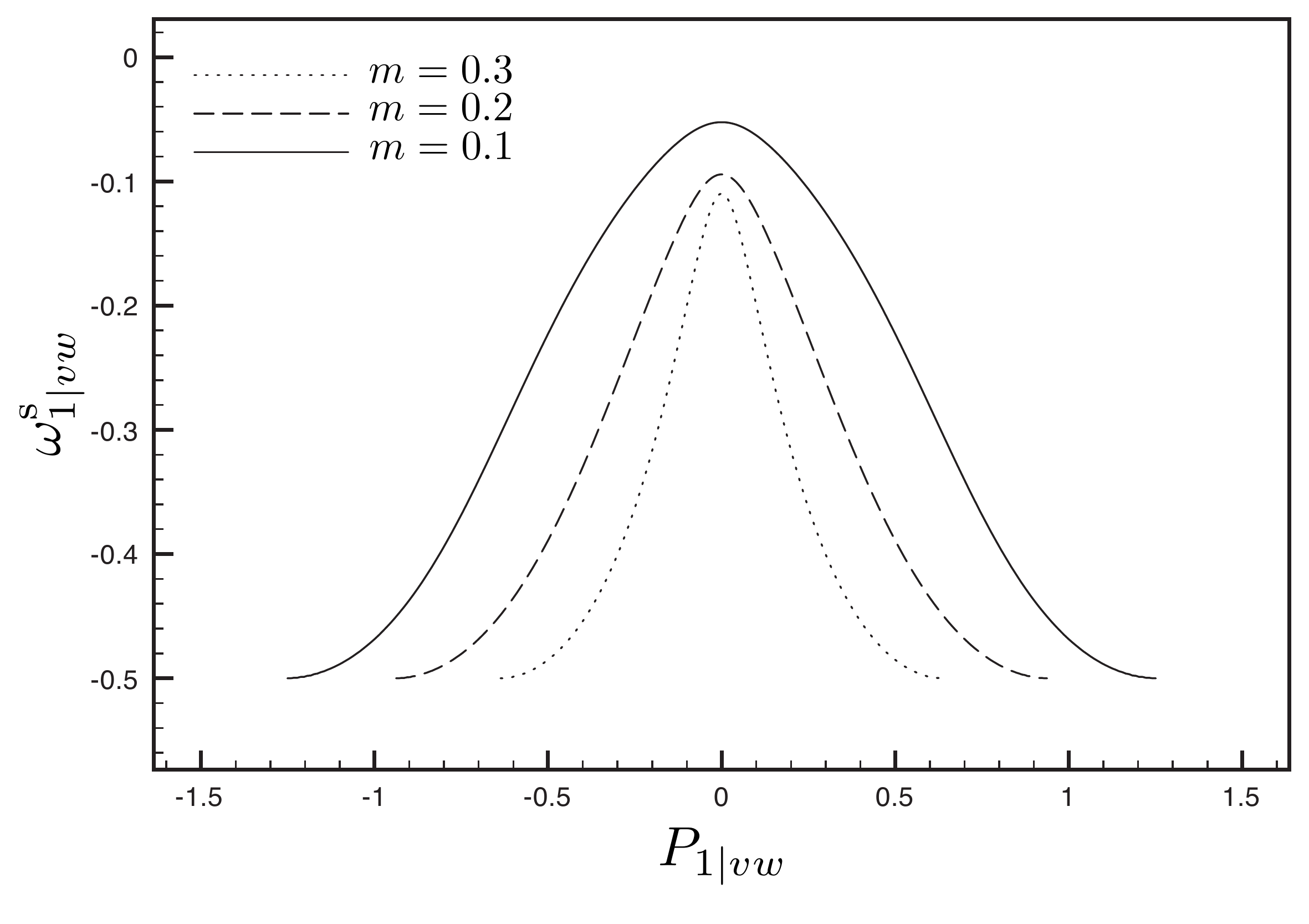}
\vspace{-0.5cm}
\caption{\small  B-model: Plots of dressed energy $\e_{1|vw}(P_{1|vw})$ and dressed charge  $\omega^{\rm s}_{1|vw}(P_{1|vw})$ at $\uh=1$ and $\mu=0$ for various values of magnetisation $m$.  }
  \la{ESvwmhf}
\end{figure}
\be
 \omega_y^{\rm s} = {1\ov2} -  \omega_{1|vw}^{\rm s}\star K_1\,,\quad
 \omega_{1|vw}^{\rm s} =   \omega_y^{\rm s}\circledast_{Q_y}K_{1} - \omega_{1|vw}^{\rm s}\star_{Q_{1|vw}}K_{1M}\,.
  \ee
Let us remark that $\omega_y^{\rm s}(\pm\infty)=1$, $\omega_{1|vw}^{\rm s}(\pm\infty)=-{1\ov2}$ and thus at zero temperature  the dressed spin jumps as soon as a magnetic field is introduced. This is true for any filling. Plots of the dressed energy and dressed spin of a $1|vw$-string are given in Figure \ref{ESvwmhf}  for various magnetisations at $\uh=1$ and $\mu=0$. The corresponding plots for $y$-particles are given in Figure \ref{ESymhf}.
\begin{figure}[t]
\includegraphics[width=0.48\linewidth]{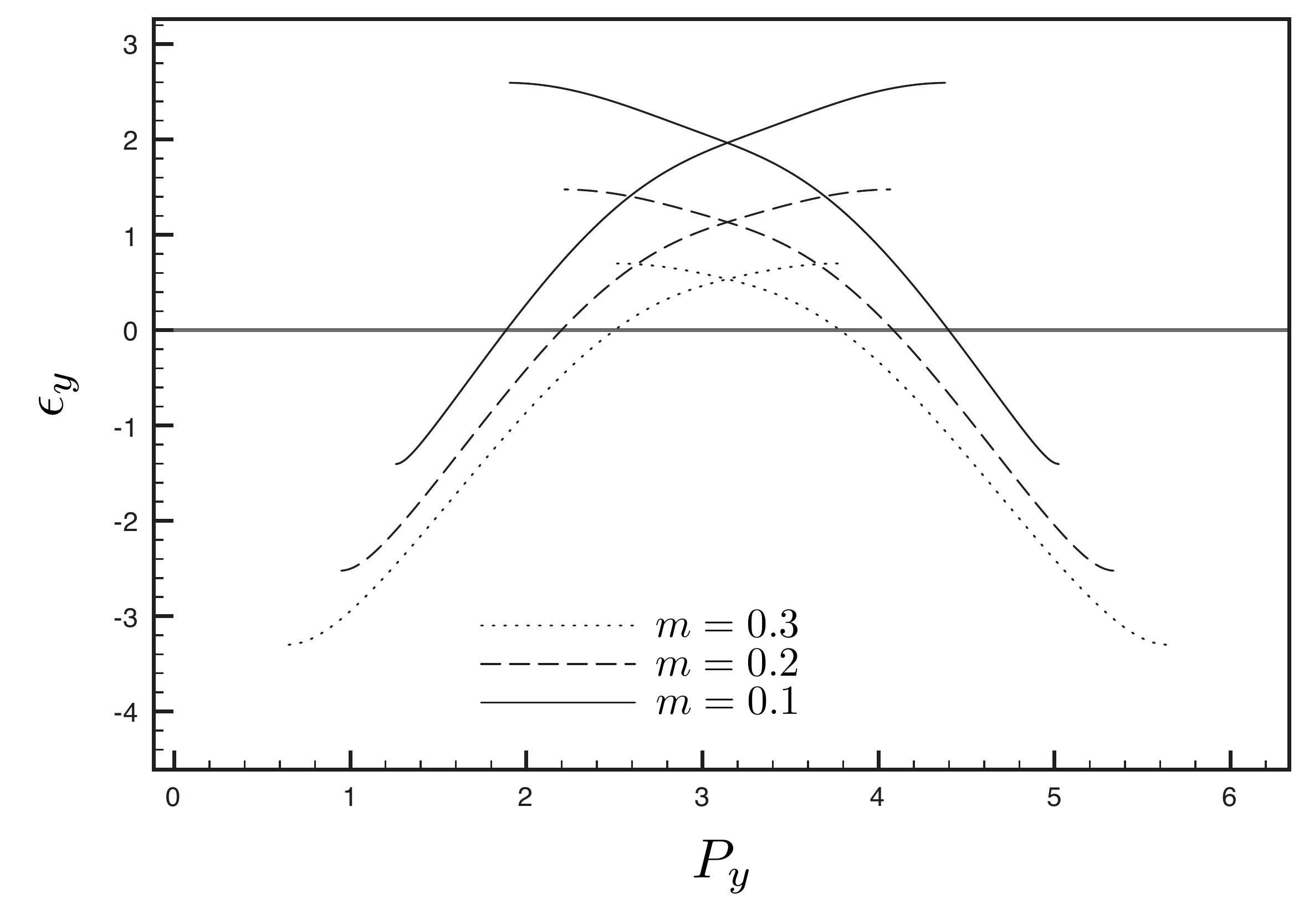}
\hfill
\includegraphics[width=0.48\linewidth]{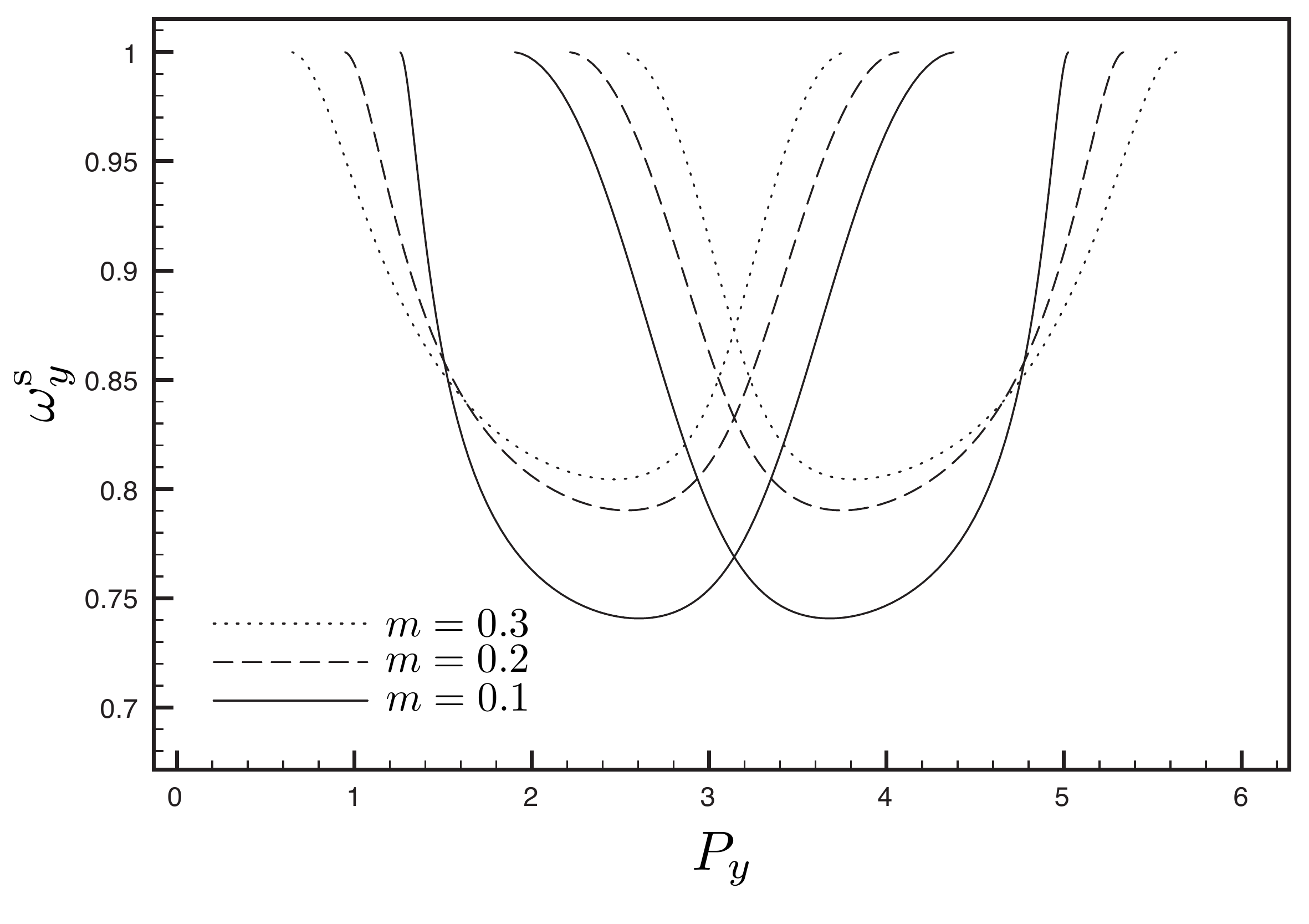}
\vspace{-0.5cm}
\caption{\small  B-model: Plots of dressed energy $\e_y(P_{y})$ and dressed charge  $\omega^{\rm s}_y(P_{y})$ at $\uh=1$ and $\mu=0$ for various values of magnetisation $m$.  }
  \la{ESymhf}
\end{figure}

 We conclude by comparing our findings to studies, respectively \cite{tJ} and \cite{EKTBA}, of related models  the supersymmetric {\it t-J} model  \cite{Lai74}-\cite{Schl87} and Essler-Korepin-Schoutens (EKS) model \cite{EKS}-\cite{EKS93}. These share similar phase diagrams, indeed in the limit of strong coupling the B-model reduces to the EKS-model with coupling $U=4$, and the supersymmetric {\it t-J} model also shares their common ground state. 
Thus the excitations are expected to be very similar and this is the case.  
The dressing of spin and charge of excitations, which we describe in detail, is  
suggested at in \cite{tJ} and is not discussed in \cite{EKTBA}. 
In particular we do not observe the existence of an electronic excitation carrying charge 1 and spin ${1\ov 2}$ for any filling as claimed in \cite{EKTBA}.
In our language the excitation they consider is understood as a $y_-$-particle and it carries these charge and spin only at zero filling where it is undressed.
An interesting feature of the EKS model is the presence of gapless excitations  for $U<4\ln2\sim2.77$ that they call localons, and in our language correspond to $M|vw$-strings with $M\geq2$. These are gapped throughout the B-model however as is to be expected due to the presence of a charge gap at half-filling.
Let us comment on a difference between the B-model and its strong coupling limit related to the hourglass-like dispersion of the $y$-particles, see Figures \ref{EPhsB} and \ref{ECyBlhf}. In the $\uh\to\infty$ limit  the dispersion curves split into two branches, one upper one lower, that touch tangentially at one point. These correspond to the $y_+$ and $y_-$ branches of the $y$-particle respectively. For the  supersymmetric {\it t-J} model excitations corresponding to the upper branch do not appear. For finite $\uh$ however the structure of the excitation is no longer of an upper and lower branch but rather of left and right moving excitations.
Finally we should comment on the advantages of our formalism over those of  say  \cite{tJ, EKTBA}. By working directly with the Bethe strings we obtain a clear description of excitations over the pseudo-vacuum reference state instead of  the somewhat unnatural reference state which is the preferred choice of \cite{tJ, EKTBA} because it makes it easier to work with Bethe roots. 
Furthermore, overcoming the need to deal directly with mode numbers and using dressing equations \eqref{eqW} to determine the dressing of spin and charge allowed us to straightforwardly identify the nature of the quasi-particle excitations.

\section*{Acknowledgements} 
The work 
of E.Q. and S.F. 
was supported by the Science Foundation Ireland under
Grant No. 09/RFP/PHY2142. 


\appendix

\renewcommand{\theequation}{\thesubsection.\arabic{equation}}

\section{Appendices}

\subsection{Conventions, definitions and notations} \la{conventions}

\subsubsection*{Convolutions}
The symbol $\star$ denotes the following ``convolution'' 
\be\la{star}
g\star h\equiv \int_{-\infty}^\infty \, dt\, g(u,t)h(t,v)\,, 
\ee
If $g$ (or $h$) is a function of a single variable then one just drops $u$ (or $v$ or both), e.g.  if $g=g(t)$ then $g\star h\equiv \int_{-\infty}^\infty \, dt\, g(t)h(t,v)$.
However, 
if  $g$ or $h$  is a kernel  defined through a function of one variable then it should be understood as $g(t,v)\equiv g(t-v)$.

In section \ref{formalism} we have expressions of the form $g_{\a\beta}\star h_{\beta\g}$ and here convolution is understood to be over the domain of the rapidity of the $\beta$-string. In section \ref{HSmods} we deal directly with the Hubbard-Shastry models for which the $y$-particles have a non-trivial domain of rapidity,  see appendix \ref{HSy}. We use the symbol  $\circledast$  to denote a contour integral in the counter-clockwise direction around the branch cut of $x(v)$, given by eqn. \eqref{xABy}.
 Explicitly, for the  A-model one has
\be\notag
g_{\a y}\circledast  h_{y \g} =\int_{|t|\le 1} \, dt\,\left( g_{\a -}(u,t)h_{-\g}(t,v) - g_{\a +}(u,t)h_{+\g}(t,v)\right) = 
g_{\a -}\hstar h_{-\g}-g_{\a+}\hstar h_{+\g}\,,~~~~~
\ee
while for the B-model
\be\notag
g_{\a y}\circledast  h_{y\g} =\int_{|t|\ge 1} \, dt\,\left( g_{\a-}(u,t)h_{-\g}(t,v) - g_{\a+}(u,t)h_{+\g}(t,v)\right)= 
g_{\a-}\cstar h_{-\g}-g_{\a+}\cstar h_{+\g}\,,~~~~~
\ee
where 
\be\notag
g_{\a\pm}(u,t)\equiv g_{\a y}(u,t\pm i0)\,,\quad h_{\pm\g}(t,v)\equiv h_{y\g}(t\pm i0,v)\,,
\ee
and $\hstar$ and $\cstar$ denote convolutions with the integration over $|t|\le1$  and $|t|\ge1$ respectively.

\subsubsection*{Matching the notations and conventions}

Most of 
our notations and conventions come from   \cite{AFS09}, and 
here we compare them to  those of \cite{bookH}.

In the Bethe ansatz we denote particles momenta as $p_j$ and auxiliary roots as $w_j$, so they are related to the ones in \cite{bookH} as $p_j \leftrightarrow k_j$, $w_j \leftrightarrow \Lambda_j$.

In the string hypothesis a $M|w$-string is a $\Lambda$ string of length $M$,
a $w$-particle is a $\Lambda$-string of length 1, 
a $M|vw$-string is a $k$-$\Lambda$ string of length $M$, and
$y$-particles could have been called $k$-particles.

In the TBA equations the Y-functions are related to the ones in \cite{bookH} as $Y_{M|w} \leftrightarrow \eta_M$, $Y_{M|vw}  \leftrightarrow \eta_M'$, $Y_{-}(\sin(k))  \leftrightarrow \zeta(k)\,,\ |k|\le\pi/2$ and $Y_{+}(\sin(k))  \leftrightarrow \zeta(k)\,,\ |k|\ge\pi/2$.

\subsubsection*{Kernels and S-matrices}

In section \ref{HSmods} we deal explicitly with various S-matrices and kernels which we list here
\begin{alignat}{2}
K_M (v) &= \frac{1}{2\pi i} \, \frac{d}{dv} \, \log S_M(v) = \frac{1}{\pi} \, \frac{\uh\, M}{ v^2+\uh^2M^2}\,,\quad S_M(v)= \frac{v - i\,\uh\, M}{v + i\,\uh\, M} \,, \la{KQkern}\\
K_{MN}(v) &= \frac{1}{2\pi i} \, \frac{d}{dv} \, \log S_{MN}(v)=K_{M+N}(v)+K_{N-M}(v)+2\sum_{j=1}^{M-1}K_{N-M+2j}(v)\,,\la{KMNkern}\\
S_{MN}(v) &=S_{M+N}(v)S_{N-M}(v)\prod_{j=1}^{M-1}S_{N-M+2j}(v)^2 =S_{NM}(v)\,,\la{SMNkern}\\
s (v) & = \frac{1}{2 \pi i} \, \frac{d}{dv} \log S(v)= {1 \ov 4\uh \cosh {\pi  v \ov 2\uh }}\,,\quad S(v)=-\tanh\big( \frac{\pi v}{4\uh} -\frac{i\pi}{4}\big)\,,
\la{skern}
\end{alignat}
Let us give also explicitly  the functions $\Theta_M(v)=2\arctan({v\ov \uh\,M})$,
\be
\Theta_{MN}(v) = \Theta_{M+N}(v)+\Theta_{N-M}(v)+2\sum_{j=1}^{M-1}\Theta_{N-M+2j}(v)\,.
\ee

\subsubsection*{Some useful identities}

\be\la{kerKid}\begin{aligned}
&1\star K_M = 1\,,\quad K_M\star K_N = K_{M+N}\,,\quad 1\star s = {1\ov 2}\,,\\
&K_1-s\star K_2 = s\,,\quad K_{M+1}-s\star K_M-s\star K_{M+2} =0\,.
\end{aligned}
\ee
\be\la{iddpE}
 {{\rm d} p_y \ov {\rm d} v}\circledast K_M=- {{\rm d} p_{M|vw} \ov {\rm d} v}\,,\quad \E_y\circledast K_M=- \E_{M|vw}\,.
\ee
\be
\pi\,\mbox{sign}\star K_M = \Theta_M\,.
\ee

\subsection{Dressed momentum as a variable}\la{DrPvar}

In the text  we have chosen to parametrise all quantities using the $v$-rapidity variable for which the S-matrices take a difference form. Here we would like to point out that 
parametrising in terms of dressed momentum has some interesting features. 
Dressed momentum as a function of $v$ is given in eq. \eqref{eqP} as $P_\a=p_\a-\rho_\beta\star\phi_{\beta\a}$. It is defined in terms of the density with respect to which the momentum gets dressed. By inversion one obtains $v_\a(P)$.

The densities parametrised in terms of $P$ are related to those parametrised in terms of $v$ as $\rho_\a(P) = \left|\Der{v_\a}{P}\right| \rho_\a\big(v_\a(P)\big)$. It is worthwhile here to choose a different normalisation for the densities so let us introduce $\tilde\rho_\a = {1\ov2\pi}\rho_\a$ and $\tilde\brho_\a = {1\ov2\pi}\brho_\a$. The equation for densities eq. \eqref{rbr} then simplies dramatically to
\be
\tilde\rho_\a(P)+\tilde\brho_\a(P)=1\,,
\ee
for the density through which the dressed momentum has been introduced,
as can be seen from \eqref{eqdP}. Remarkably the densities are non-interacting in this parametrisation.
Moreover the densities are related to the functions $Y_\a$ simply as
\be
\tilde\rho_\a(P) = {1\ov 1+Y_\a(P)} \,.
\ee

Next consider the expansion \eqref{defxi} in terms of dressed momentum. Let us define $\tilde\zeta_\a$ through
\be\la{defxiP}
\tilde P_{\a,k} - P_{\a,k} ={\tilde\zeta_{\a}(P_{\a,k})\ov L}\,.
\ee
The closed equation for $\tilde\zeta_\a$
takes the form
\be\la{xirP}
\tilde\zeta_\a=\tilde\zeta_\beta\tilde\rho_{\beta}\star \tilde K_{\beta\a}-\p_{\a{\Tadd}}+\p_{\a{\Trem}} \,,
\ee
where $ \tilde K_{\a\beta}(P,Q) = \Der{v_\a(P)}{P} K_{\a\beta}(v_\a(P)-v_\beta(Q))$, and here the shift function is just $-\tilde \zeta_\a$. The dressing equations also take a nice form, for example for dressed charge one has
\be
\omega_\a = \barew_\a + \omega_\beta \tilde\rho_\beta \star \tilde K_{\beta\a}\,.
\ee
Finally, the induced charge term \eqref{indW} becomes $\Delta W_{\rm ind}=-\tilde\zeta_\a \omega_\a\star \tilde\rho_{\a}'$, with $\tilde\rho_{\a}'=\Der{\tilde\rho_{\a}}{P}$.


\subsection{The $y$-particles of the Hubbard-Shastry models} \la{HSy}

In this appendix we describe the $y$-particles of the Hubbard-Shastry models. These do not fit into the classification of strings as type 1 and type 2 in section \ref{formalism} and so we indicate how the formalism developed there can be applied to them. First it is convenient  to introduce  the two functions
\be\la{xABy}
x_{\rm A}(v)=v+v\sqrt{1-\frac{1}{v^2}}\,,\quad x_{\rm B}(v)=v+i\sqrt{1-v^2}\,.
\ee
The function $x_{\rm A}$ has a cut along $(-1,1)$ and the function  $x_{\rm B}$ has a cut along $(-\infty,-1)\cup (1,\infty)$. Let us define $\I^{\rm A, B}_+$ and $\I^{\rm A, B}_-$ to be respectively the upper and lower edges these cuts. For values of $v$ on the cuts we define $x_{\rm A, B}(v)=x_{\rm A, B}(v+i0)$. Then, since $v={1\ov 2}(y+1/y)$, for any given $v$ one has two possible corresponding $y$-roots
and they can be parametrized as
\be\la{yr}
y_+=x(v)\,,\quad y_-={1\ov x(v)}\,, 
\ee
so that in terms of the $v$-rapidity the $y$-roots are split into two subsets. 
The $y$-particles can thus be  regarded as  a string whose $v$-rapidity takes values on a two-sheeted Riemann surface. For the  Hubbard  and A-models we take $x(v) = x_{\rm A}(v)$ and so for these models $\I_y=\I^{\rm A}_+\cup\I^{\rm A}_-$, while for the B-model we take $x(v) = x_{\rm B}(v)$ and here $\I_y=\I^{\rm B}_+\cup\I^{\rm B}_-$.

Now let us address the definition of the counting function for these $y$-particles.  As for the type 1 and type 2 strings identified in section \ref{formalism}, we  wish to define the counting function $z_y$ so that it is continuous everywhere except at the maximum of the pseudo-energy $\e_y$ of the $y$-string. This is achieved at the cost of the counting function being an increasing function of $v$, and thus care must be taken about how the densities are defined in order to ensure that they are positive. 
Let us restrict ourselves to the parity invariant Hubbard-Shastry models and 
let us assume without loss of generality\footnote{This is the case for both the A- and B-models. The alternative possibility applies to the Hubbard model and proceeds in exactly the same manner under interchange of $+$ and $-$.} that the minimum of $\e_y(v)$ is on the edge $\I^{\rm A,B}_+$.
Then we define the counting function as
\be\la{zay}
L \,\s_+ z_y(v)=\pi{\varphi_y}+ {L\, p_y(v)}+\sum_\beta\sum_{n=1}^{N_\beta}\,\p_{y\beta}(v,v_{\beta,n})\,,
\ee
where the range of $p$ is chosen to be continuous everywhere along $\I^{\rm A,B}_\pm$ except at the value $v=v_{\rm max}\in\I^{\rm A,B}_-$, the value of $v$ corresponding to the maximum of $\e_y(v)$, and the scattering phases $\p_{y\beta}$ are defined as in eq. \eqref{case1} for the case $y_+=x_{\rm A}$ and as in eq. \eqref{case2} for the case $y_+=x_{\rm B}$. 
For the Hubbard-Shasrty models $\s_+=\mbox{sign}(\Der{p_+}{v})=-1$ and $\s_-=\mbox{sign}(\Der{p_-}{v})=1$ and so $z_y$ is an increasing function of $v$ for $v\in \I^{\rm A,B}_+$ and it is a decreasing  function of $v$ for $v\in \I^{\rm A,B}_-$.  This ensures that the counting function is increasing on the contour  clockwise around the cut of $x_{\rm A,B}(v)$ (that goes along $\I^{\rm A,B}_+$ on the upper side and oppositely along $\I^{\rm A,B}_-$ on the other) with the exception of the point $v_{\rm max}\in\I^{\rm A,B}_-$ at which it is discontinuous. When introducing the densities however it is necessary to include the factor $\s_-$ for the density of roots on the edge  $\I^{\rm A,B}_-$ in order to ensure that this density is positive. Thus we have
\be
\rho_{\pm} + {\bar{\rho}}_{\pm} = \frac{1}{2 \pi} \left|\Der{p_\pm}{v}\right|+ K_{\pm\beta}\star\rho_{\beta} \,,
\ee
where 
\be
K_{\pm\beta}=\s_\pm \cK_{\pm\beta} \,,\quad \cK_{\pm\beta}(v)={1\ov 2\pi i}{d\ov dv}\log S_{\pm\beta}(v) \,,
\ee
and
\be
\rho_y(v) =\begin{cases}  \rho_+(v) &\mbox{if } v\in \I_+ \\
\rho_-(v) &\mbox{if } v\in \I_- \end{cases}\,,\quad \brho_y(v) = \begin{cases}  \brho_+(v) &\mbox{if } v\in \I_+ \\
\brho_-(v) &\mbox{if } v\in \I_- \end{cases}\,.
\ee
Finally let us remark that $K_{+\beta}=-K_{-\beta}$ as $S_{+\beta}=S_{-\beta}$, and so 
\be
k_{y\beta}=1\circledast K_{y\beta} =0\,,
\ee
for all $\beta$-strings.



\end{document}